%% file: jetprops.tex
\documentclass[reprint, amsmath, amssymb, aps, prd,showpacs,twocolumn,groupedaddress]{revtex4-1}
\pdfoutput=1 
\usepackage{graphicx} 
\usepackage{dcolumn}
\usepackage{bm}
\usepackage{setspace}
\usepackage{epsfig}
\usepackage{epstopdf}
\usepackage{multirow}
\usepackage{url}

\usepackage{xspace}
\usepackage{booktabs, threeparttable}
\usepackage{rotating}
\usepackage{tabularx}
\usepackage{array}
\newcommand\T{\rule{0pt}{2.6ex}}
\newcommand\B{\rule[-1.2ex]{0pt}{0pt}}
\usepackage{bigstrut}
\newcommand{\pT}{\ensuremath{p_{T}}\xspace}
\newcommand{\antikt}{anti-$k_{t}$\xspace}

\newcommand{\figref}[1]{Fig.~\ref{fig:#1}}
\newcommand{\eref}[1]{Eq.~\ref{eq:#1}}
\newcommand{\tabref}[1]{Table~\ref{tab:#1}}
\newcommand{\secref}[1]{Sec.~\ref{sec:#1}}

\newcommand{\avg}[1]{\left< #1 \right>} % for e.g. <M>
\def\pythia{{\sc Pythia}}

\def\herwigpp{{\sc Herwig++}}

\def\powheg{{\sc PowHeg}}

\def\perugia{{\sc Perugia$2010$}}
\def\ambt{{\sc AMBT1}}
\def\aue{{\sc AUET2B}}
\def\hepdata{{\sc HepData}}
\def\rivet{{\sc Rivet}}

\usepackage[hyperindex, breaklinks]{hyperref}
\hypersetup{linkcolor=blue,citecolor=blue,filecolor=black,urlcolor=blue} 

\usepackage{preprintcover,atlasphysics}

\PreprintCoverPaperTitle{ATLAS measurements of the properties of jets for boosted particle searches}  % paper title
\PreprintIdNumber{CERN-PH-EP-2012-149}  % CERN preprint number
\PreprintCoverAbstract{
Measurements are presented of the properties of high transverse momentum jets, produced in proton-proton
collisions at a center-of-mass energy of $\sqrt{s} =$ 7 TeV. The data correspond to an integrated luminosity of
$35$ \ipb\ and were collected with the ATLAS detector in 2010.
Jet mass, width, eccentricity, planar flow and angularity are
measured for jets reconstructed using the \antikt algorithm with distance
parameters $R =$ 0.6 and 1.0, with transverse momentum \pT\ $>$ 300
GeV and pseudorapidity $|\eta| <$ 2.
The measurements are compared to the expectations of Monte Carlo generators
that match leading-logarithmic parton showers to leading-order, or
next-to-leading-order, matrix elements. The generators describe the general features of the jets, although discrepancies are observed in some distributions.
}  
\PreprintJournalName{Phys. Rev. D}  % journal name                                                              

\begin{document}
\title{ATLAS measurements of the properties of jets for boosted particle searches}
\author{The ATLAS Collaboration}

\begin{abstract}
Measurements are presented of the properties of high transverse momentum jets, produced in proton-proton
collisions at a center-of-mass energy of $\sqrt{s} =$ 7 TeV. The data correspond to an integrated luminosity of
$35$ \ipb\ and were collected with the ATLAS detector in 2010.
Jet mass, width, eccentricity, planar flow and angularity are
measured for jets reconstructed using the \antikt algorithm with distance
parameters $R =$ 0.6 and 1.0, with transverse momentum \pT\ $>$ 300
GeV and pseudorapidity $|\eta| <$ 2.
The measurements are compared to the expectations of Monte Carlo generators
that match leading-logarithmic parton showers to leading-order, or
next-to-leading-order, matrix elements. The generators describe the general features of the jets, although discrepancies are observed in some distributions.
\end{abstract}

\maketitle

\section{Introduction\label{sec:introduction}}

The high center-of-mass energy at the Large Hadron Collider (LHC) combined with the 
coverage and granularity of the ATLAS calorimeter 
provide an excellent environment to study hadronic jets. Measurements
of dijet cross sections~\cite{xsec, cms_dijetxsec}, jet
shapes~\cite{shapes, cms_shapes}, jet substructure~\cite{substructpaper} and angular 
correlations~\cite{angular, cms_ang}
have already been published using the data taken by the ATLAS and CMS
Collaborations in 2010.

Massive, hadronically decaying particles produced with a
significant boost (such as top quarks, Higgs
bosons, or new particles) will tend to have collimated
decays, such that their decay products are contained within a single
jet. The substructure of jets
resulting from such decays is expected to result in deviations in the
observables measured here for light-quarks and gluons, thus providing
discriminating power in heavy particle searches.

The observable jet properties presented here are mass, width, eccentricity, planar
flow and angularity. All of these have been shown to be useful in
Monte Carlo studies in the search for high transverse momentum (\pT),
massive
particles~\cite{boost_procs,bdrs,Almeida:2008yp,wtag,Chekanov:2010vc,CDFmass,ttH},
and together they provide an important set of probes of the
substructure of jets. 

Three
of these (mass, planar flow and angularity) have recently been
measured by CDF~\cite{CDFmass} at the Tevatron. 
Angularities are a family of infrared-safe quantities that have characteristic
distributions for two-body decays, while planar flow discriminates between
two-body and many-body decays and, for large jet masses (above about 100~GeV), is
largely independent of the jet mass. Eccentricity is a complementary
observable to planar flow, with which it is highly anti-correlated.
Jet width is a dimensionless quantity related to the jet mass and
is thus expected to retain much of the discriminatory power without
being as sensitive to the 
detector effects on energy scale and resolution that can hinder a mass
measurement. 

Jet substructure measurements can be particularly
vulnerable to `pileup', i.e. particles produced in multiple $pp$
interactions that occur in addition to the primary interaction, within the sensitive time of the detector. These additional interactions result in diffuse, usually soft,
energy deposits throughout the central region of the detector -- the region of
interest for the study of high \pT\ jets. This additional energy
deposition can be characterized by the number of
reconstructed primary
vertices ($N_{\rm{PV}}$)~\cite{pv1,pv}, with events having a single good vertex ($N_{\rm{PV}}$~=~1) being considered free from the effects of pileup. The 2010 ATLAS data set provides a unique opportunity to study
these effects; a significant fraction of the 2010 data set
comprises $N_{\rm{PV}}$~=~1 events, making this
data set ideal for evaluating the effects of pileup on jet substructure
measurements. This data set has an average $N_{\rm{PV}} \simeq$ 2.2.

\section{The ATLAS detector}

 The ATLAS detector~\cite{ATLAS:detector} at the LHC was
 designed to study a wide range of physics. It covers
 almost the entire solid angle around the collision point with layers of tracking
 detectors, calorimeters and muon chambers. Tracks and vertices are reconstructed with the  inner detector, which
 consists of a silicon pixel detector, a silicon strip detector and a
 transition radiation tracker, all  immersed in a 2 T axial magnetic field provided
 by a superconducting solenoid.

 The ATLAS reference system is a Cartesian right-handed co-ordinate system,
 with the nominal collision point at the origin.
 The anti-clockwise beam direction defines the positive $z$-axis,
 while the positive $x$-axis is defined as pointing from the collision
 point to the center of the LHC ring and the
 positive $y$-axis points upwards. The azimuthal angle $\phi$ is measured around the beam
 axis, and the polar angle $\theta$ is the angle measured with respect
 to the $z$-axis. The pseudorapidity
 is given by $\eta=-\ln\tan(\theta/2)$. Transverse momentum is defined
 relative to the beam axis.

 For the measurements
 presented here, the high-granularity calorimeter systems are
 of particular importance. The ATLAS calorimeter system provides fine-grained
 measurements of shower
 energy depositions over a large range in $\eta$.

 Electromagnetic calorimetry in the range $|\eta| < $ 4.9 is
 provided by liquid-argon (LAr) sampling calorimeters. This calorimeter system
 enables measurements of the shower energy in up to four depth
 segments. For the jets measured here, the transverse
 granularity ranges from 0.003$\times$0.10 to
 0.10$\times$0.10 in $\Delta\eta \times \Delta\phi$, depending on depth segment and pseudorapidity.

 Hadronic calorimetry
 in the range $|\eta| < $ 1.7 is provided by a steel/scintillator-tile
 sampling calorimeter.
 This system enables measurements of the shower energy deposition
 in three depth segments at a transverse granularity of typically 0.1$\times$0.1.
 In the end-caps ($|\eta| >$ 1.5), LAr technology is used for the
 hadronic calorimeters that match the outer $\eta$-limits of the end-cap
 electromagnetic calorimeters. This system enables four measurements in depth
 of the shower energy deposition at a transverse granularity of either
 0.1$\times$0.1 (1.5 $< |\eta| <$ 2.5) or 0.2$\times$0.2 (2.5 $< |\eta|
 <$ 3.2).

\section{Monte Carlo simulation}

The QCD predictions for the hadronic final state
in inelastic $pp$ collisions are based on
several Monte Carlo generators with different tunes. %The tunes are based on the
%parameter sets provided by the Professor Collaboration~\cite{proftunes}.

The \pythia\ 6.423 generator~\cite{Sjostrand:2006za} with the ATLAS Minimum Bias Tune 1 (\ambt)~\cite{ambtref} parameter set is
used as the primary generator for comparisons with the data and for extracting
corrections to the data for detector effects. 
The \ambt\ tune uses the MRST
LO$^{*}$~\cite{Sherstnev:2007nd} parton distribution function (PDF)
set with leading-order (LO) perturbative QCD matrix elements for
2$\rightarrow$2 processes and a leading-logarithmic, \pT-ordered parton shower
followed by fragmentation
into final-state particles using a string model
~\cite{Andersson:1983ia} with Lund functions~\cite{Lund} for light quarks and Bowler functions~\cite{Bowler} for
heavy quarks. In addition to charged particle measurements from ATLAS minimum bias
data~\cite{MB900, MB7}, the \ambt\ tune uses data from LEP, SPS and the Tevatron.

An additional \pythia\ tune, \perugia~\cite{Skands:2010ak,tunes},
is used for comparison with \ambt. The \perugia\ tune also
uses data from LEP, SPS and the Tevatron and additionally improves the
description of jet shape measurements in LEP data.
The CTEQ5L~\cite{Pumplin:2002vw} PDF set is used. This tune of
\pythia\ is used in the calculation of the systematic uncertainties on
the measurements and for comparison with the data, along with the
\herwigpp\ 2.4.2 generator with its default settings~\cite{Bahr:2008pv}. 

The more recent \herwigpp\ 2.5.1 generator is included for comparison
with the final measurements at particle level, as are the \aue\ tune~\cite{aue, aue2}
of \pythia\ 6.423, the
\powheg\ generator interfaced to this same \pythia\ tune, and the
\pythia\ 8.153 generator~\cite{pythia8} with tune
4C~\cite{tunes,aue}. The major difference between \herwigpp\ versions 2.4.2 and 2.5.1 is the
inclusion of color-reconnections in the latter. The \powheg\
generator, which implements next-to-leading-order (NLO) calculations within a shower Monte
Carlo context~\cite{powheg, powheg2, powheg3, powheg4}, uses the
CTEQ6M~\cite{Pumplin:2002vw} PDF set.

Generated events are passed through the ATLAS detector simulation
program~\cite{soft},
which is based on {\sc Geant4}~\cite{Agostinelli2003250}. The quark--gluon string model with an additional precompound~\cite{QGSP} is used for the
fragmentation of nuclei, and the Bertini cascade
model~\cite{BERT} is used to describe the interactions of hadrons with
the nuclear medium.

Monte Carlo events are reconstructed and analyzed using the same event
selection and simulated trigger as for the data.
The size and position of the collision beam spot and
the detailed description of detector conditions during the data-taking runs are included
in the simulation.

\section{Event selection}
Events containing pileup can be identified by
the presence of more than one primary vertex in the event,
herein referred to as $N_{\rm{PV}}$~$>$~1. Events recorded in the 2010 ATLAS
data set contain an average $N_{\rm{PV}} \simeq$~2.2 and include a significant
fraction of $N_{\rm{PV}}$~=~1 events ($\simeq$~28$\%$); these may be used for
testing the pileup correction methods. 

After applying data-quality
requirements, the data sample corresponds to a total integrated
luminosity of 35.0 $\pm$
1.1~\ipb~\cite{LUMI, LUMI2}.

\subsection{Trigger selection}
Events must pass the ATLAS first-level trigger requiring a jet (built from calorimeter towers 
with a granularity of $0.1\times0.1$ in $\Delta \eta \times \Delta \phi$) with transverse
energy $E_{\mathrm{T}}\geq$
95~GeV. 
The selection efficiency of this
trigger has been found to be
close to $100\%$ for events satisfying the offline selection criteria
implemented here, with a negligible dependence on jet mass~\cite{substructpaper}.

\subsection{Primary vertex selection}
All events are required to have at least one good primary
vertex. This is defined as a vertex with at least five tracks
with \pT\ $>$ 150 MeV and both transverse and longitudinal impact parameters
consistent with the LHC beamspot~\cite{pv1,pv}. The analysis presented here makes use of the full 2010 data set. The
requirement of $N_{\rm{PV}}$~=~1 is applied only
where derivation of pileup corrections is not possible.

\subsection{High \pT\ jet selection}
Jets are reconstructed from locally calibrated topological
clusters~\cite{clustering} using the \antikt algorithm~\cite{antikt} with distance
parameters of $R =$ 0.6 and 1.0.  Jets satisfying \pT\ $>$ 300~GeV and $|\eta| <$ 2 are selected for
analysis. 
Any event containing an $R =$ 0.6 jet with \pT\
$>$ 30~GeV that fails to satisfy the criteria~\cite{JEM} designed to
safeguard against jets caused by transient detector effects and beam
backgrounds is
excluded from this analysis.

In simulated data, jets are reconstructed from locally calibrated
topological clusters to derive corrections for pileup and determine the systematic uncertainties and detector correction factors. The corrected data
distributions are then compared to Monte Carlo predictions at
particle level; in this case jets are reconstructed from stable
particles as opposed to clusters. Particles are deemed to be stable for the purpose of jet reconstruction
if their mean lifetimes are longer than 10~ps. Neutrinos and muons are
excluded, just as they are for the Monte Carlo-based jet energy scale
calibration that is applied to the data. This exclusion has a negligible effect
on the final measurements. 

The total numbers of jets in data satisfying the selection criteria detailed
here are $\sim$122,000 $R =$ 1.0 jets and $\sim$87,000 $R =$ 0.6 jets; however, only
the highest \pT\ jet in each event is selected for this analysis.
The total numbers
selected for analysis are $\sim$83,000 $R
= $ 1.0 jets and $\sim$62,000 $R =$ 0.6 jets. 

\section{Substructure observables and their correlations \label{sec:observables}}

\subsection{Jet mass \label{sec:obsmass}}

The jet mass $M$ is calculated from the energies and momenta of its
constituents (particles or clusters) as follows:

\begin{equation}\label{eq:jetmass}
M^{2}=\left(\sum_{i}E_{i}\right)^{2}-\left(\sum_{i}\vec{p}_{i}\right)^{2}, 
\end{equation}

\noindent 
where $E_i$\ and $\vec{p}_i$\ are the energy and three-momentum of the
$i^{th}$ constituent. The sum is over all jet constituents in this
and all subsequent summations. The standard ATLAS reconstruction
procedure is followed: clusters have their masses set to zero,
while Monte Carlo particles are assigned their correct masses.

\subsection{Jet width \label{sec:obswidth}}
The jet width $W$ is defined as:

\begin{equation}\label{eq:width} 
W=\frac{\sum_i \Delta R^{i}\: p_{\mathrm{T}}^{i}} {\sum_i p_{\mathrm{T}}^{i}}, 
\end{equation}
\noindent
where $\Delta R^i \:$=$\:\sqrt{(\Delta \phi_i)^2 + (\Delta \eta_i)^2}$ is the radial distance between the jet axis 
and the $i^{th}$ jet constituent and $p_{\mathrm{T}}^{i}$ is the constituent
\pT\ with respect to the beam axis.

\subsection{Eccentricity \label{sec:obsecc}}

The jet eccentricity $\mathcal{E}$ is calculated using a principal
component analysis (PCA)~\cite{Chekanov:2010vc}. The PCA method
provides the vector which best describes the energy-weighted
geometrical distribution of the jet constituents in the ($\eta, \phi$) plane.
The eccentricity is used to characterize the deviation of the jet profile from a
perfect circle in this plane, and is defined as
\begin{equation}
\mathcal{E} = 1 - \frac{v_{\mathrm{min}}}{v_{\mathrm{max}}},
\end{equation}
\noindent
where $v_{\mathrm{max}}$ ($v_{\mathrm{min}}$) is the maximum (minimum) value of
variance of the jet constituents' positions with respect to the
principal vector. The calculation consists
of the following steps:

\begin{enumerate}
\item For each jet the energy-weighted centers in $\eta$ and $\phi$ are
calculated as:

\begin{equation}  \label{eq:eccstep1} 
\bar{\phi}_{\mathrm{jet}}=\frac{ \sum_i\Delta \phi_i \: E_i}{ \sum_i E_i}, \qquad 
\bar{\eta}_{\mathrm{jet}}=\frac{ \sum_i \Delta \eta_i \: E_i}{ \sum_i  E_i}, 
\end{equation}
\noindent
where the energy and position in the ($\eta, \phi$) plane of the $i^{th}$ constituent with respect to
the jet axis are denoted by $E_i$, $\Delta \eta_{i}$ and
$\Delta \phi_{i}$.

\item The PCA is performed to determine the vector $\vec{x}_1$ in
   $(\eta,\phi)$ space that passes through the energy-weighted center
   of the face of the jet and results in a minimum in the variance of
   the constituents' positions. The angle $\theta$ of this vector with respect to the
   jet center ($\bar{\eta}_{\mathrm{jet}}, \bar{\phi}_{\mathrm{jet}}$) is given by:

\begin{equation} \label{eq:rotangle}
\tan 2\theta  =
\frac{ 2 \times \sum_i E_i \Delta\phi_i \Delta \eta_i}{ \sum_i E_i (\Delta
  \phi_i^{2} - \Delta\eta_i^2)} 
\end{equation}

and the angle of the orthogonal vector $\vec{x}_2$ is $\theta-\frac{\pi}{2}$.

\item The energy-weighted variances $v_1$ and $v_2$ with respect to $\vec{x}_1$ and $\vec{x}_2$ are calculated as:

\begin{flalign}
v_1 = \frac{1}{N} \sum_{i} E_{i}\>
(\: \cos \theta \Delta\eta_{i} - \sin \theta \Delta\phi_i \: )^2, \nonumber \\ 
v_2 = \frac{1}{N} \sum_{i} E_{i}\>
(\: \sin \theta \Delta\eta_{i} + \cos \theta \Delta\phi_{i} \: )^2, 
\end{flalign}

\noindent
where $N$ is the number of constituents.

\item Finally, the largest value of the variance is assigned to  $v_{\mathrm{max}}$ and
the smallest to $v_{\mathrm{min}}$.
The jet eccentricity ranges
from zero for perfectly circular jets to one for jets that appear pencil-like in the ($\eta,\phi$) plane.
\end{enumerate}

Eccentricity is measured for jets in the mass range $M
>$ 100~GeV; this is the mass region of
interest for the search for a Higgs boson or other massive, hadronically decaying particles predicted in various
extensions to the Standard Model. 

\subsection{Planar flow \label{sec:obspf}}

A variable complementary to the eccentricity is planar flow $P$~\cite{PhysRevD.79.074012,Almeida:2008yp,thaler_wang}. The planar flow measures
the degree to which the jet's energy is evenly spread over the plane across
the face of the jet (high planar flow) versus spread linearly across
the face of the jet (small planar flow).

To calculate planar flow, 
one first constructs a two-dimensional matrix $I_{E}^{kl}$:

\begin{equation}\label{eq:pf1}
I_{E}^{kl} = \frac{1}{M}\sum_{i} \frac{1}{E_{i}} p_{i,k} p_{i,l}.
\end{equation}

Here, $M$ is the jet mass, $E_{i}$ is the energy of the $i^{th}$
constituent of the jet and $p_{i,k}$ and $p_{i,l}$ are the $k$ and $l$
components of its transverse momentum calculated with respect to the jet axis. The planar flow is: 

\begin{equation}\label{eq:pf2}
P = 4\times \frac{\mathrm{det}(I_{E})}{\mathrm{Tr}(I_{E})^{2}}. 
\end{equation}

Vanishing or low planar flow corresponds to a linear energy deposition, as in 
the case of a two-pronged decay, while completely 
isotropic energy distributions are characterized by unit planar flow ~\cite{Almeida:2008yp}. Jets with many-body kinematics
 are expected to have a planar flow distribution that peaks towards unity. 
In general, QCD jets have a rising $P$ distribution that peaks at 
$P\:$=$\:$ 1; the hadronization process has contributions from many 
soft gluons and is largely isotropic. However, jets with high \pT\ and high mass are well-described by a single hard gluon emission.
Consequently, these jets have a planar flow distribution
that peaks at a low value~\cite{CDFmass}. 
The planar flow distributions
are measured in the context of boosted, massive particle searches by applying a
mass cut, 130 $< M <$ 210~GeV, consistent with the window in which one
would expect to observe a boosted top quark decay collimated within a
single jet. The contribution from top quark decays in this subset of
the data is
negligible -- here we measure the properties of light-quark and gluon jets
that constitute a substantial fraction of the background in boosted top quark measurements.

\subsection{Angularity \label{sec:obsang}}
Angularities ($\tau_{a}$) are a family of observables that are sensitive to the degree of symmetry in the energy flow inside a jet. 
The general formula for angularity~\cite{Almeida:2008yp} is given by:

\begin{equation}\label{eq:angul}
\tau_{a} = \frac{1}{M}\sum_{i}{E_{i}\mathrm{sin}^{a}\theta_{i}[1-\mathrm{cos}\theta_{i}]^{1-a}},
\end{equation}

\noindent
Here $a$ is a parameter that can be chosen to emphasize radiation
near the edges ($a < 0$) or core ($a > 0$) of the jet, $M$ is the jet mass, $E_{i}$ is the energy of the $i^{th}$ jet constituent and $\theta_{i}$ is its angle with respect to the jet axis. In the limit of small-angle radiation ($\theta_{i} \muchless 1$), $\tau_{a}$ is approximated by:

\begin{equation}\label{eq:angul2}
\tau_{a} \simeq \frac{2^{(a-1)}}{M}\sum_{i}{E_{i}\theta_{i}^{(2-a)}}.
\end{equation}
\noindent

Angularities are infrared-safe for $a \leq 2$~\cite{CDFmass}. In the
analysis presented here, \eref{angul} with a value of $a=-2$ is
used. The $\tau_{-2}$ observable can be used as a discriminator for distinguishing QCD jets from boosted particle decays by virtue of the broader tail expected in the QCD distribution \cite{Almeida:2008yp}. 
At a given high mass, the angularity of jets with two-body kinematics should peak around a minimum value 
$\tau_{a}^{\mathrm{peak}}\simeq(\frac{M}{2p_{\mathrm{T}}})^{1-a}$, which corresponds to
the two hard constituents being in a symmetric \pT\
 configuration around the jet axis. 
An estimate for the 
maximum of the distribution can also be calculated in the limit of
small angle radiation, $\tau_{a}^{\mathrm{max}} \simeq
(\frac{2}{R})^{a}(\frac{M}{2p_{\mathrm{T}}})$~\cite{CDFmass}, which
corresponds to a hard constituent close to the jet axis and a soft constituent on the jet edge.  

The measurement of $\tau_{-2}$ is aimed primarily at
testing QCD, which makes predictions for the shape of the $\tau_{-2}$
distribution in jets where the small angle approximation is valid. For
this reason, this measurement is made only for \antikt\ jets with $R
=$ 0.6.

Here, $\tau_{-2}$ is measured for jets in the mass range
100 $< M <$ 130~GeV. This mass region
is chosen to have minimal contributions from 
hadronically decaying $W$ or $Z$ bosons or boosted top quarks (\pythia\ predictions estimate
a relative fraction below 0.2$\%$). 

\subsection{Correlations between the observables\label{sec:correlations}}

The levels of correlation between the variables presented here provide information that is valuable in deciding which variables may
potentially be used together in a search for boosted
particles. Here the correlation factors between pairs of variables are
calculated as their covariance divided by the product of their
standard deviations:

\begin{equation}\label{cf}
\rho = \frac{cov(x,y)}{\sigma_{x} \sigma_{y}}.
\end{equation}

\begin{figure}
\begin{center}
\includegraphics[width=0.44\textwidth,angle=0]{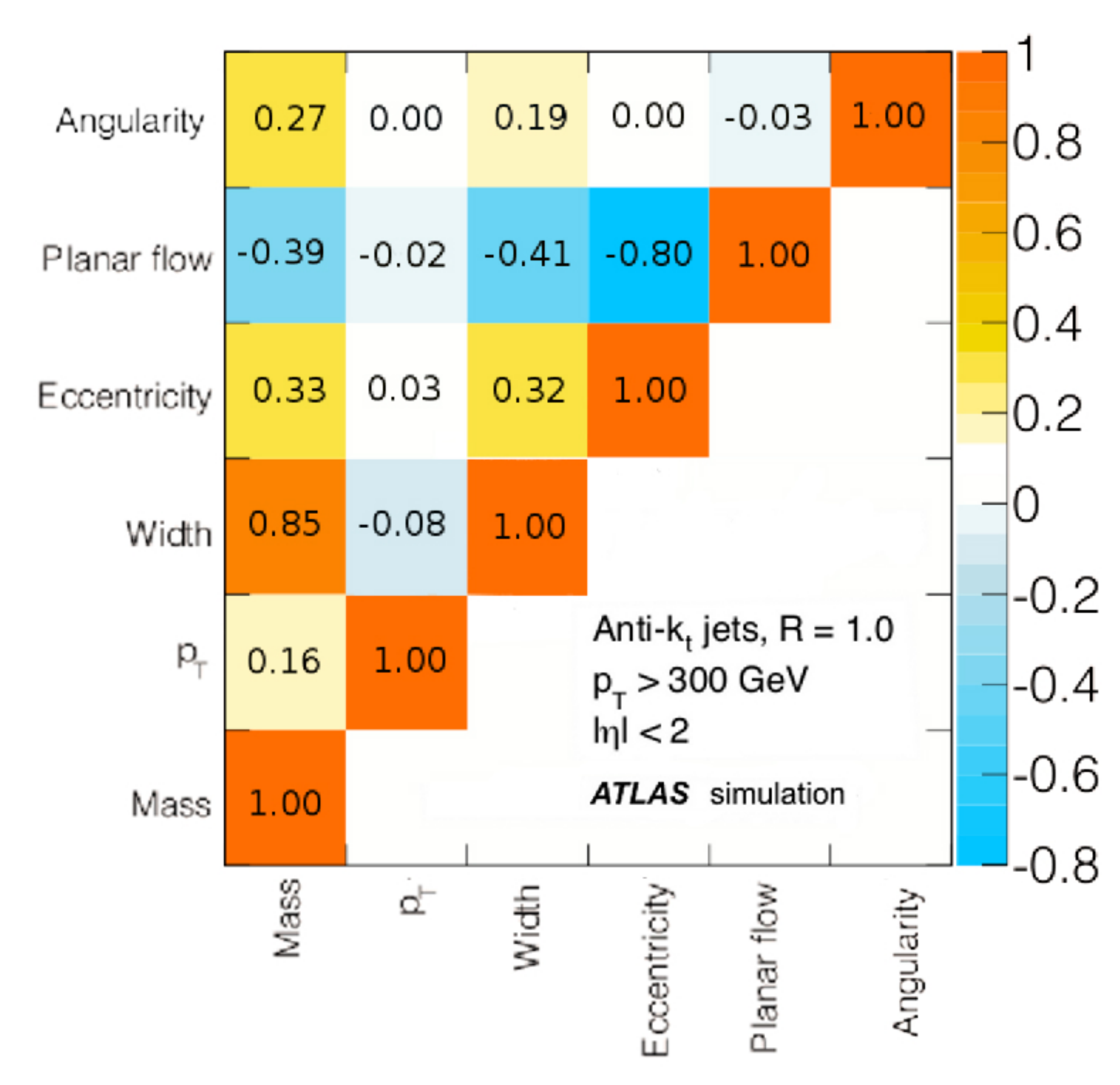}
\includegraphics[width=0.44\textwidth, angle=0]{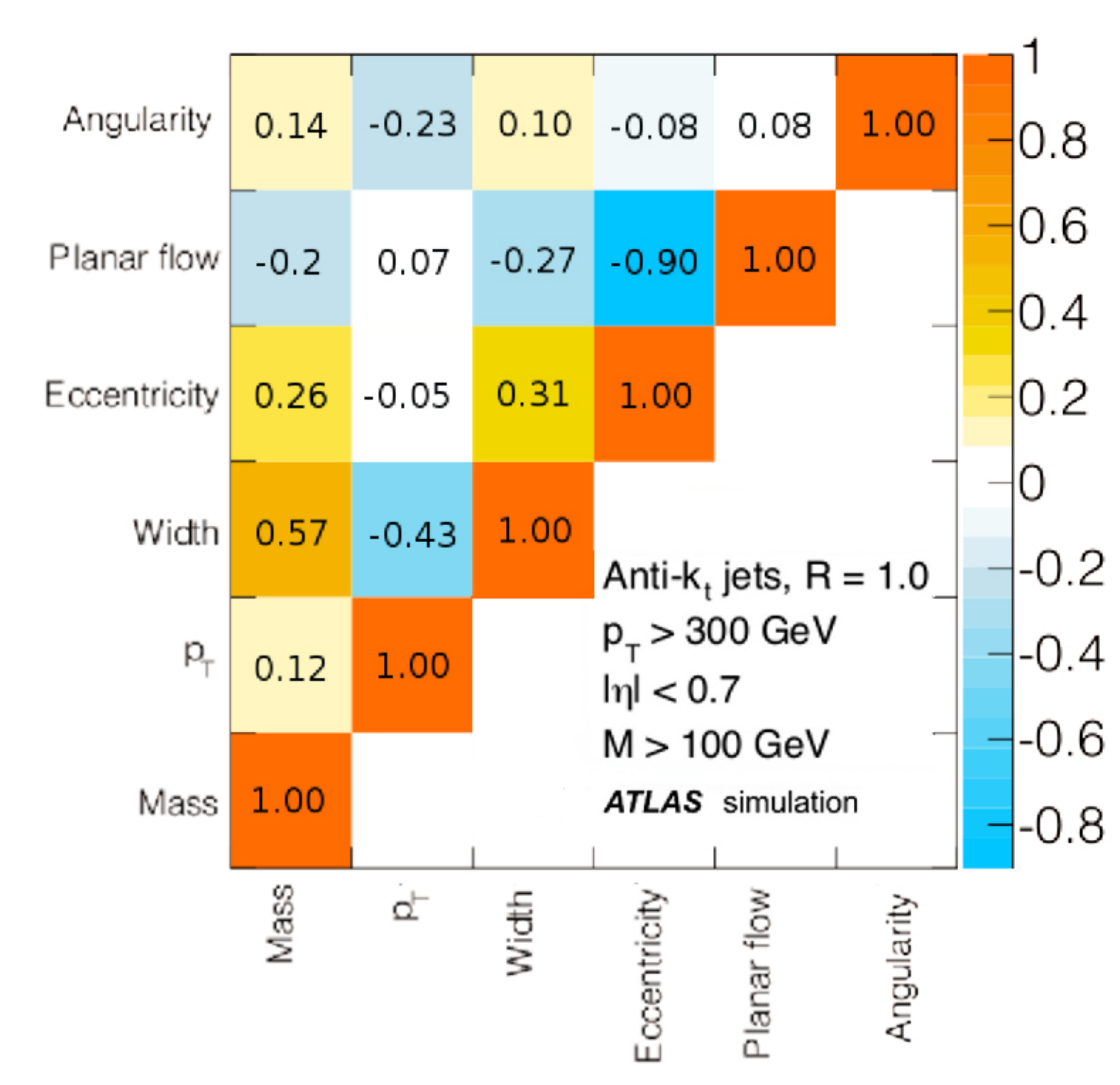}
\end{center}
\caption{ The correlation coefficients between pairs of variables
 calculated in \pythia\ at particle level for $R =$ 1.0 jets with no mass constraint
 (top) and with a mass constraint of $M >$ 100~GeV (bottom).}
\label{fig:correlations}
\end{figure}

Summaries of the correlations between all of the observables
studied here are shown in \figref{correlations} for $R=$1.0 jets
in \pythia\ at particle level. The
coefficients are shown both with and without a jet mass cut of $M>$
100~GeV -- the individual mass constraints for each observable are
dropped here to allow the correlations between them to be calculated. Jets subjected to a mass cut are also restricted to $|\eta|<$ 0.7. This additional restriction on $\eta$ is
applied wherever a mass cut is made on the observables presented here;
this has a negligible effect on the shapes of the distributions while allowing direct
comparisons with other measurements of the same quantities~\cite{CDFmass}.

The strongest correlations observed are those between jet mass and width (85$\%$) and between planar flow and
eccentricity (--80$\%$). The correlation between mass and width
reduces considerably when jets are required to be in the kinematic
region $M >$~100~GeV. This trend is followed by almost all observables. The planar flow
and eccentricity, however, are even
more strongly anti-correlated in high mass jets (--90$\%$).
The correlation between mass and \pT\ is weak (12--16$\%$).
Angularity is largely uncorrelated with all of the other observables.

\section{Corrections for pileup and detector effects}\label{sec:corrections}

 The contribution from pileup is
  measured using the complementary cone method  first introduced by the CDF experiment ~\cite{CDFmass, MIPhen}. A complementary
cone is drawn at a right-angle in azimuth to the jet
($\phi_{\mathrm{comp}}=\phi_{\mathrm{jet}} \pm \frac{\pi}{2}$, $\eta_{\mathrm{comp}}=\eta_{\mathrm{jet}}$)
and the energy deposits in this cone are added into
the jet such that the effect on each of the jet properties can be
quantified.
The shift in each observable after this addition is attributed to
pileup and the underlying event (UE), the latter being the diffuse radiation present in all events and partially coherent with the hard scatter. The effects of these two sources are separated by comparing events with $N_{\rm{PV}}$~=~1 (UE only) to those with $N_{\rm{PV}}$~$>$~1 (UE and pileup): the difference between the average shift for single-vertex and multiple-vertex events is attributed to the contribution from pileup only. 

The presence of additional energy in events with $N_{\rm{PV}}$~$>$~1 affects the
substructure observables in different ways; the effect of pileup on
the shape of the $\tau_{-2}$ distribution is negligible (below ~1$\%$) in
this data set, and so no corrections are applied. The other
observables under study have their distributions noticeably
distorted by the presence of pileup. The \pT-dependent corrections for this effect are
applied to the mass, width and eccentricity distributions, whilst the planar flow distribution of high mass jets is
measured only in events with $N_{\rm{PV}}$~=~1. There are a small number of jets ($\sim$~100
\antikt $R =$ 0.6 jets)  in the high mass range ($M >$ 130~GeV), making it too difficult to derive robust pileup
corrections for planar flow (which is limited to this mass range in
this analysis) in this data set.

\subsection{Pileup corrections for $R =$ 0.6 jets\label{sec:pileupfit}}

\begin{figure}
\begin{center}
\includegraphics[scale=0.31, angle=0]{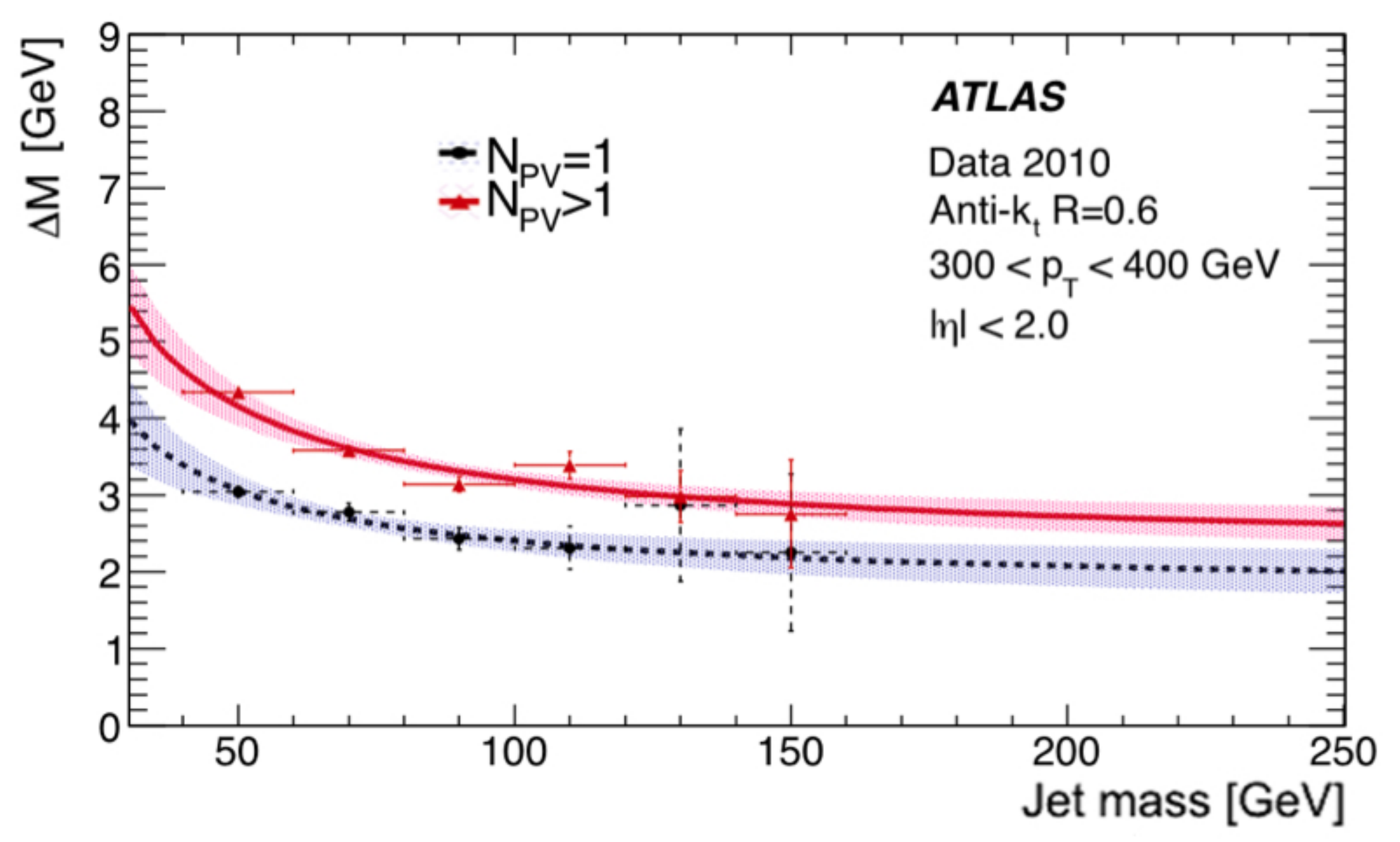}
\end{center}
\caption{ The size of the mass shift in \antikt $R
 =$ 0.6 jets with 300 $< \pT\ <$ 400~GeV in jets with pileup and UE
 ($N_{\rm{PV}}$~$>$~1, average $N_{\rm{PV}} \simeq$ 2.2) and with UE alone ($N_{\rm{PV}}$~=~1). The curves are fits of the form $\Delta M = p_{0_M} + \frac{p_{1_M}}{M}$. The difference between the curves gives the contribution to
the jet mass from pileup only. The $1\sigma$ uncertainties on the fits are shown in the error bands.}
\label{fig:mass_ue_mpi}
\end{figure}

The mass shift due to the UE and pileup in $N_{\rm{PV}}$~=~1 and $N_{\rm{PV}}$~$>$~1 events is shown in
\figref{mass_ue_mpi} for $R =$ 0.6 jets in the range 300 $<$ \pT\ $<$ 400~GeV. The shift follows the expected behavior, given
by:

\begin{equation}
\Delta M = p_{0_M} + \frac{p_{1_M}}{M},
\end{equation}
\noindent
where $p_{i_M}$ and their associated uncertainties are determined from
the data. $\Delta M$ is the increase in the jet mass due to the addition of the
energy deposits in the complementary cone to the jet. Corrections to the jet
mass are limited to the region  $M>$ 30~GeV, as the
$\frac{p_{1_M}}{M}$ parameterization uses a leading-order
approximation~\cite{MIPhen} and is only valid for $\Delta M \muchless M$. This has a
negligible effect on the final measurements in the range
$M >$ 20~GeV, as illustrated for $R =$ 1.0 jets in \figref{PUperf}; low mass,
high \pT\ jets tend to have a small contribution from pileup. 

The corresponding parameterizations for the shifts in width $\Delta W$
and eccentricity $\Delta \mathcal{E}$ are:

\begin{flalign}
&\Delta W = p_{0_W} + p_{1_W}W ,\nonumber \\
&\Delta \mathcal{E} = p_{0_\mathcal{E}} + p_{1_\mathcal{E}} \mathcal{E} + p_{2_\mathcal{E}} \mathcal{E}^{2}.
\end{flalign}
\noindent
The pileup corrections for width and eccentricity are
applied to jets across the full mass range.

\subsection{Pileup corrections for $R =$ 1.0 jets\label{sec:pileupscaling}}

The complementary cone technique cannot be applied directly to $R =$ 1.0 jets due to the high probability of overlap between the complementary
cone and the jet; scaling factors are therefore applied to the corrections measured for $R =$ 0.6 jets. 

The scaling behaviour is determined experimentally by comparing the
pileup-dependence in $R =$ 0.6 and $R =$ 0.4 jets. For each observable,
the shifts for $R =$ 0.4 jets are fit to a functional form. The shifts
for $R =$ 0.6 jets are then fit to a scaled version of this function,
where all parameters are fixed at their $R =$ 0.4 values and the
scaling is the only free parameter of the fit. The measured $R$-dependence is
then validated with a comparison between $R =$ 1.0 jets in $N_{\rm{PV}}$~$>$~1
and $N_{\rm{PV}}$~$=$~1 events.

The predicted (observed) behaviors for the scaling of the shifts in mass
and width are:

\begin{equation}\label{eq:dmD}
\Delta M: p_{i_{M}} \sim R^{4}\; (R^{3.5}),
\end{equation}

\begin{equation}\label{eq:dwD}
 \Delta W:  p_{0_W} \sim R^{3}\; (R^{2.5}),\;\; p_{1_W} \sim R^{2}\; (R^{1}).
\end{equation}

\noindent
The phenomenological predictions~\cite{SalamNonPert} for scaling are
used, and the discrepancies between predictions and observations are considered systematic uncertainties in this procedure.
%Here, the scaling of the shift in width is different for the two
%coefficients in the parameterization; this is not the case for the
%scaling of mass or eccentricity.

There is no phenomenological
prediction for the scaling of $\Delta \mathcal{E}$ with pileup,
therefore the nominal value of the scaling of the shift in this variable is measured in
data. The measurements find
the scaling of the parameterization to be a function of mass:

\begin{eqnarray}
 \Delta \mathcal{E}(M < 40~ \rm{GeV})&: &p_{i_\mathcal{E}} \sim R^2, \nonumber\\
 \Delta \mathcal{E}(M \geq 40~ \rm{GeV})&: &p_{i_\mathcal{E}} \sim R^3.
\end{eqnarray}

\noindent
The measured scaling is varied between  $R^2$ and $R^3$ across the mass
range in order to determine a conservative estimate of the systematic
uncertainty introduced by this procedure.

\begin{figure}
\begin{center}
\includegraphics[scale=0.46, angle=0]{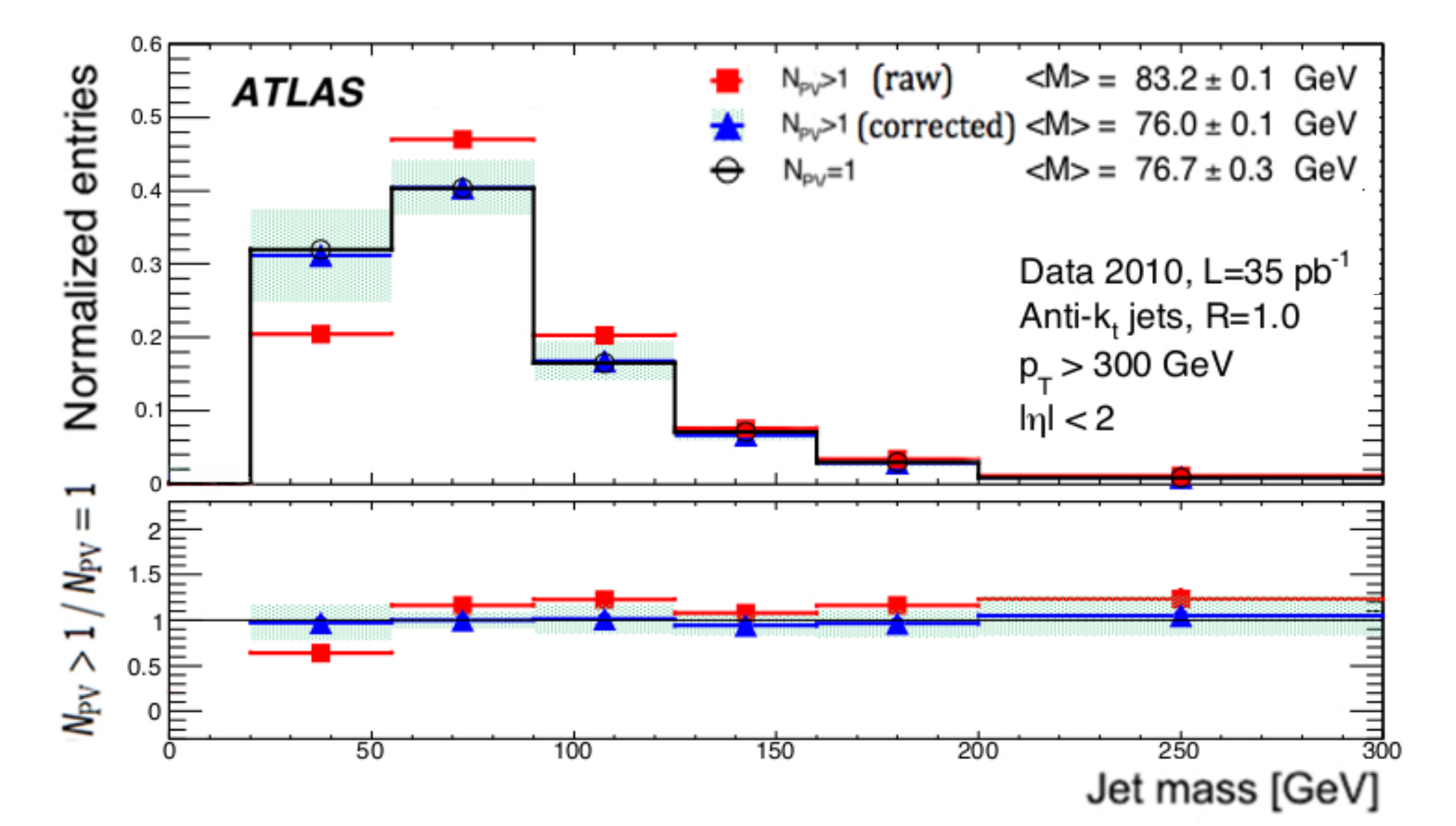}
\includegraphics[scale=0.46, angle=0]{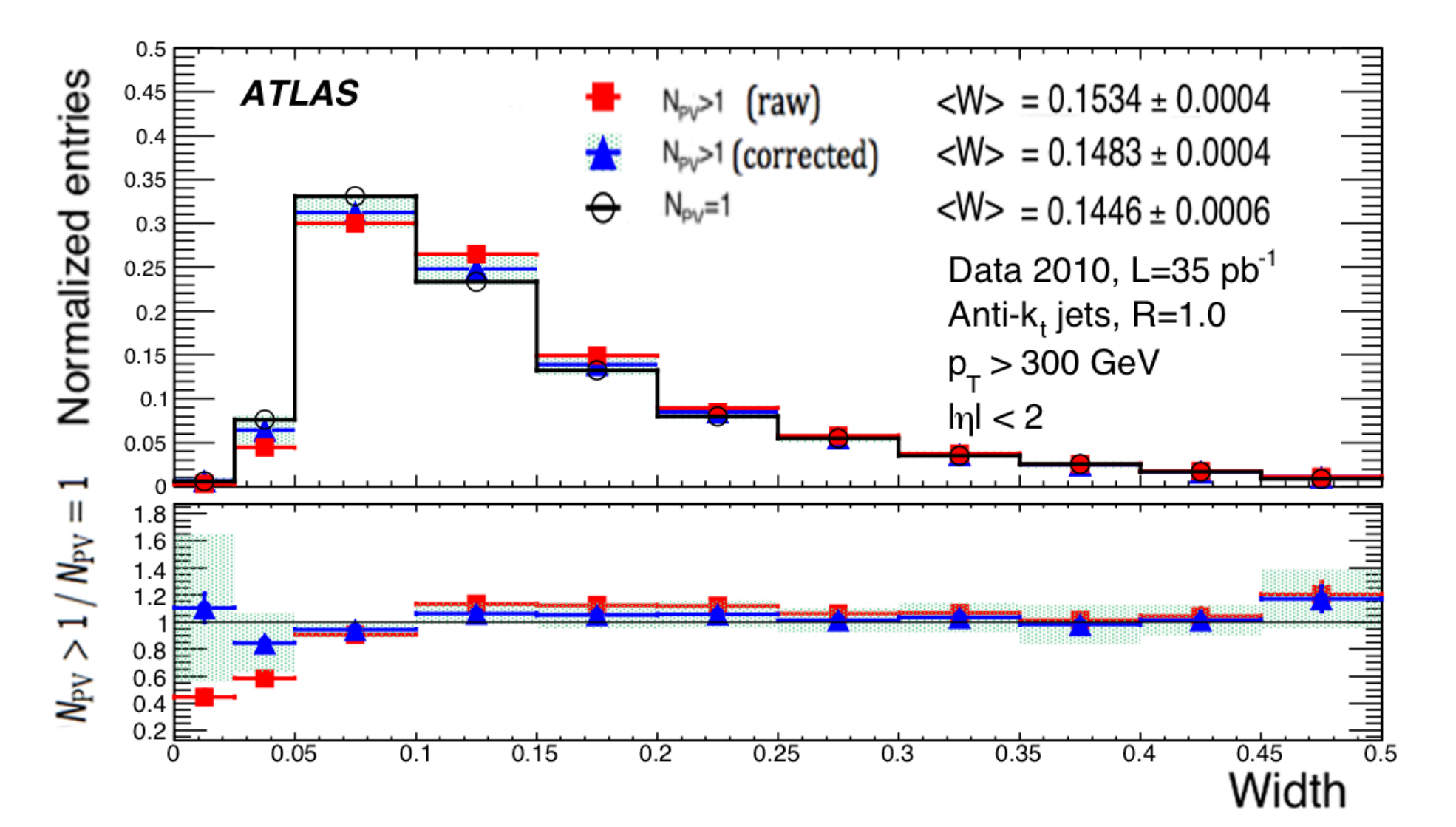}
\includegraphics[scale=0.46, angle=0]{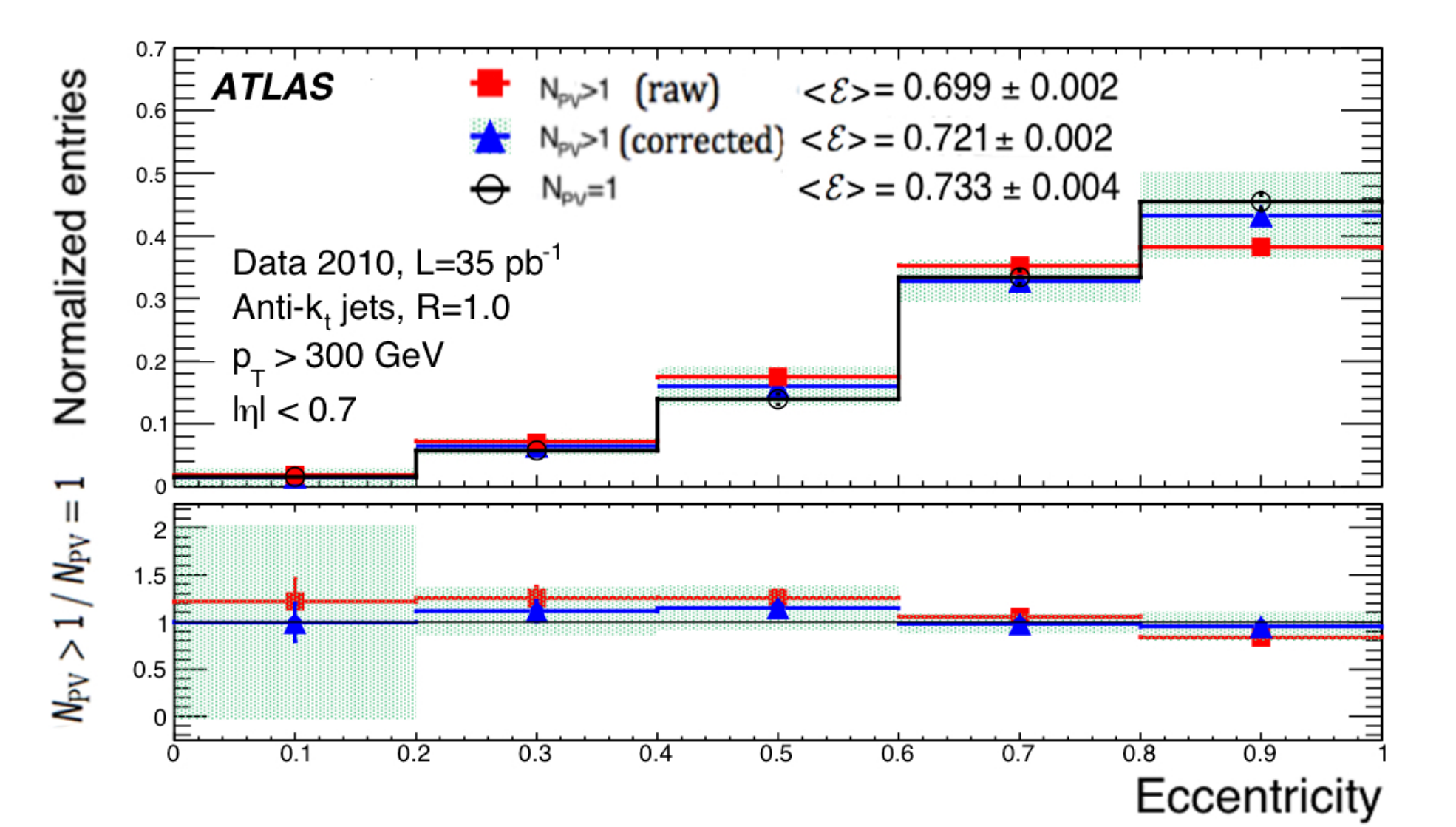}
\end{center}
\caption{
The mass, width and eccentricity distributions before and after the pileup corrections. The (red) squares indicate the uncorrected data in
the full data set, the (black) circles indicate the subset of
this data with $N_{\rm{PV}}$~=~1 and the (blue) triangles indicate the full
data set after pileup corrections. The mean value of each distribution
is indicated in the legend with the corresponding statistical uncertainty. The lower region of each figure shows
the measured ratio of $N_{\rm{PV}}$~$>$~1 to $N_{\rm{PV}}$~=~1 events.}
\label{fig:PUperf}
\end{figure}

The performance of the pileup correction procedure in the case of
mass, width and eccentricity is
shown in \figref{PUperf}. The observable most sensitive to pileup is the jet mass; the
mean $R =$ 1.0 jet mass is shifted upwards by $\sim$ 7~GeV in events with $N_{\rm{PV}}$~$>$~1, and there
is a significant change in the shape of the mass distribution. In the case of jet width and eccentricity, the effect of pileup is a
small ($\sim$ 5$\%$) shift towards wider, less eccentric jets.
This supports the expected behavior: width is less sensitive to
pileup than mass, making it a promising alternative to mass as a
criterion for selecting jets of interest in boosted particle searches in the high pileup
conditions of later LHC operations.
For all observables the discrepancies between the pileup-corrected distributions and those for
events with $N_{\rm{PV}}$~=~1 are small, and agreement is obtained within the
systematic uncertainties on the corrections. 

\subsection{Corrections for detector effects}

After correcting the distributions for pileup, each distribution is
corrected to particle level, using bin-by-bin corrections for detector effects. 
The bin migrations due to detector effects are determined and controlled by increasing the bin sizes until all 
bins have a purity and efficiency above 50$\%$ according to Monte Carlo predictions, where purity and efficiency are defined as:
      \begin{equation}\label{eq:pur}
      p_{i}=\frac{\mathcal{A}_{i}^{\mathrm{part+det}}}{\mathcal{A}_{i}^{\mathrm{det}}},\; \;
      e_{i}=\frac{\mathcal{A}_{i}^{\mathrm{part+det}}}{\mathcal{A}_{i}^{\mathrm{part}}}.
      \end{equation}

Here $\mathcal{A}_{i}^{\mathrm{part+det}}$ is the number of
detector-level jets (reconstructed from locally calibrated clusters) in bin $i$ that have a particle-level jet (reconstructed
from stable Monte Carlo particles), matched
within $\Delta R <$ 0.2 and falling in the same bin. $\mathcal{A}_{i}^{\mathrm{part}}$ is the total
number of particle-level jets in bin $i$ and $\mathcal{A}_{i}^{\mathrm{det}}$ is the total
number of detector-level jets in bin $i$.

The particle-level value for an observable in bin $i$ is found by multiplying
its measured value by the relevant correction factor $C_i$:

\begin{equation}\label{eq:corfacs}
C_i =\frac{\mathcal{A}_{i}^{\mathrm{part}}}{\mathcal{A}_{i}^{\mathrm{det}}}.
\end{equation}

The size of the corrections varies quite significantly
between observables and between bins, being  around 20$\%$ for mass,
(5--10$\%$ around the
peak, 20$\%$ elsewhere) and width (30$\%$ in the peak for $R =$ 0.6
jets, 1--5$\%$ elsewhere). The corrections for eccentricity are below 10$\%$ in the peak, increasing to 40$\%$ in the most sparsely
populated bin. The detector corrections for angularity and planar flow
are smaller, generally around 0--5$\%$.

\section{Systematic uncertainties}

The experimental systematic uncertainties can be
divided into three categories: how
well-modeled the observables are in Monte Carlo simulations
(\secref{sys_mc}), the modeling of the detector material and cluster reconstruction (\secref{sys_det}) and the pileup corrections
(\secref{sys_pu}). These are evaluated by determining the difference
in the factors obtained after the application of systematic variations
to the samples used in the correction for detector effects. The
dominant sources of uncertainty, described in detail below, arise from varying
the cluster energy scale (CES) and from the differences found when
the calculation of detector corrections is done using the \herwigpp\
Monte Carlo sample in place of \pythia\ \ambt. These dominant
effects are shown in \figref{sys6det} for $R =$ 0.6 jets and
\figref{sys10det} for $R =$ 1.0 jets.

\subsection{Uncertainties on the Monte Carlo model \label{sec:sys_mc}}

The distributions are corrected to particle level using the correction factors $C_{i}$ determined with a specific Monte Carlo
generator, inclusive of parton shower, hadronization and UE model, which in this
case is \pythia\ with the \ambt\ tune. To determine the uncertainty introduced
on the final measurement by choosing this particular model to
calculate the detector correction factors, the differences in these $
C_{i}$ are found when the \pythia\ \ambt\ tune is replaced with the
\perugia\ tune, and with \herwigpp\ (2.4.2).

\begin{figure*}
\begin{center}
\includegraphics[scale=0.8,angle=0]{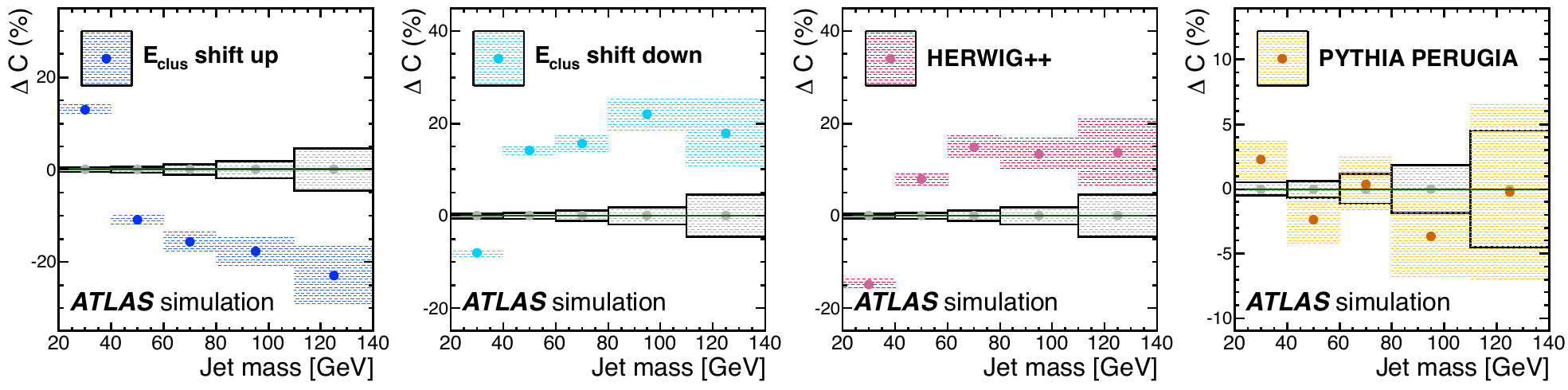}
\includegraphics[scale=0.8,angle=0]{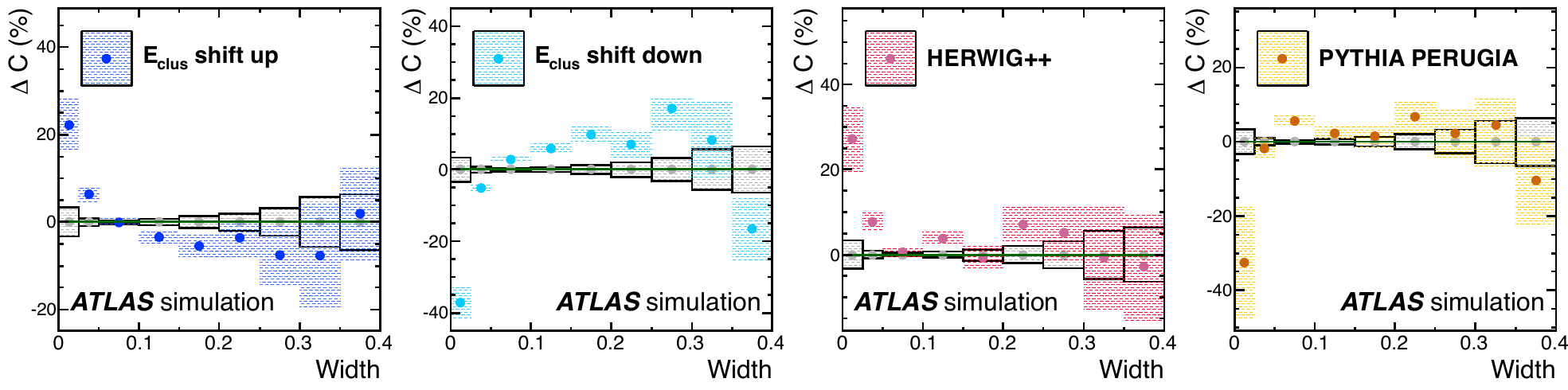}
\includegraphics[scale=0.8,angle=0]{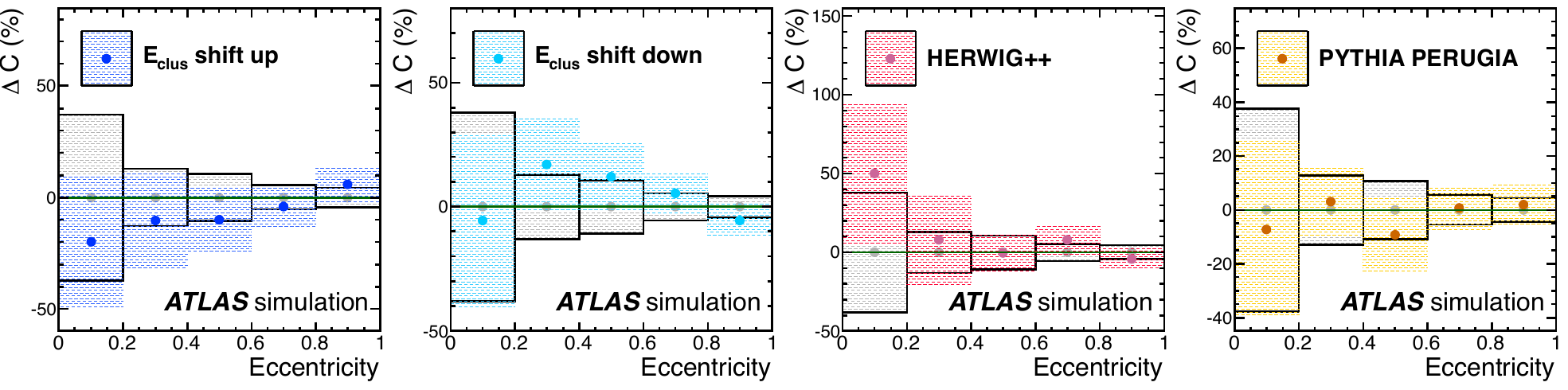}
\includegraphics[scale=0.8,angle=0]{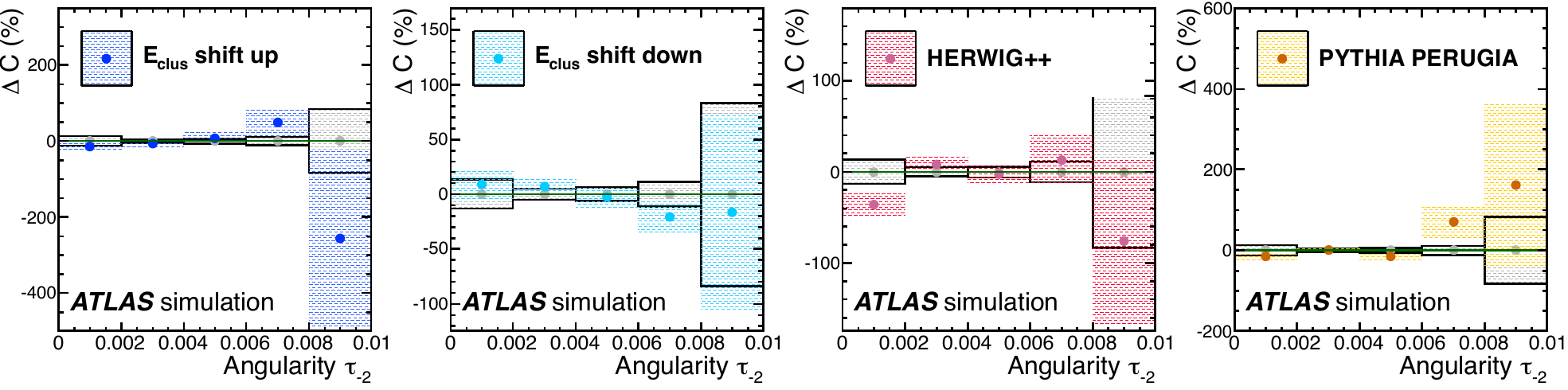}
\end{center}
\caption{The dominant sources of systematic uncertainty on the
  measurements are those resulting in large variations in the
  detector correction factors $C$. These correction factors are found bin-by-bin using
  $R =$ 0.6 jets in a \pythia\ \ambt\ sample with upward and downward variations of the cluster energy
scale (first and second columns), and by using \herwigpp\ (third
column) and \pythia\ \perugia\ (fourth column) in place of \pythia\ \ambt. The differences $\Delta C$ found
when comparing the correction factors obtained with the
baseline \pythia\ \ambt\ sample are shown here for
each of the properties measured in $R =$ 0.6 jets. The shaded bands
indicate the statistical uncertainties.}
\label{fig:sys6det}
\end{figure*}

\begin{figure*}
\begin{center}
\includegraphics[scale=0.8,angle=0]{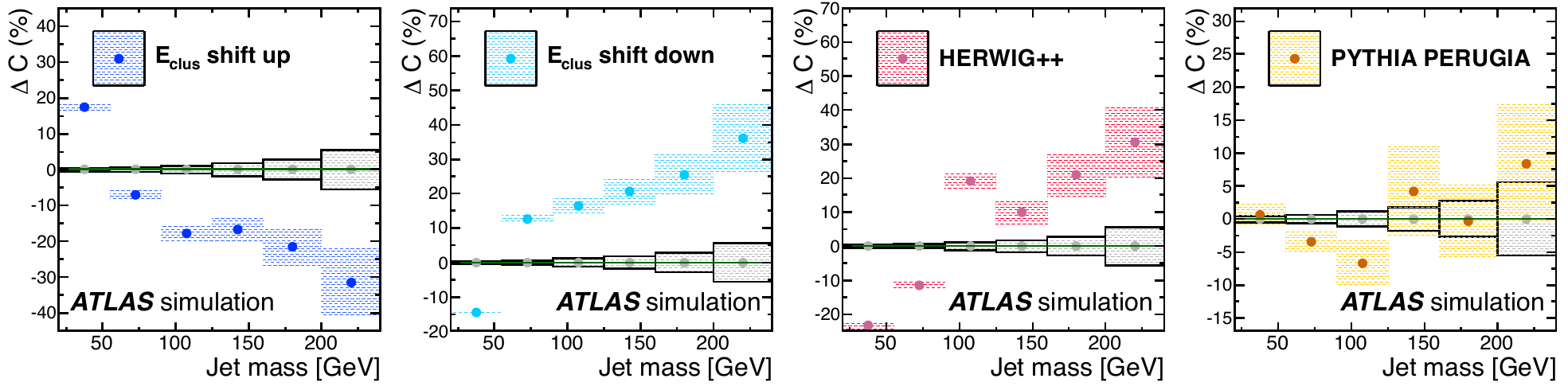}
\includegraphics[scale=0.8,angle=0]{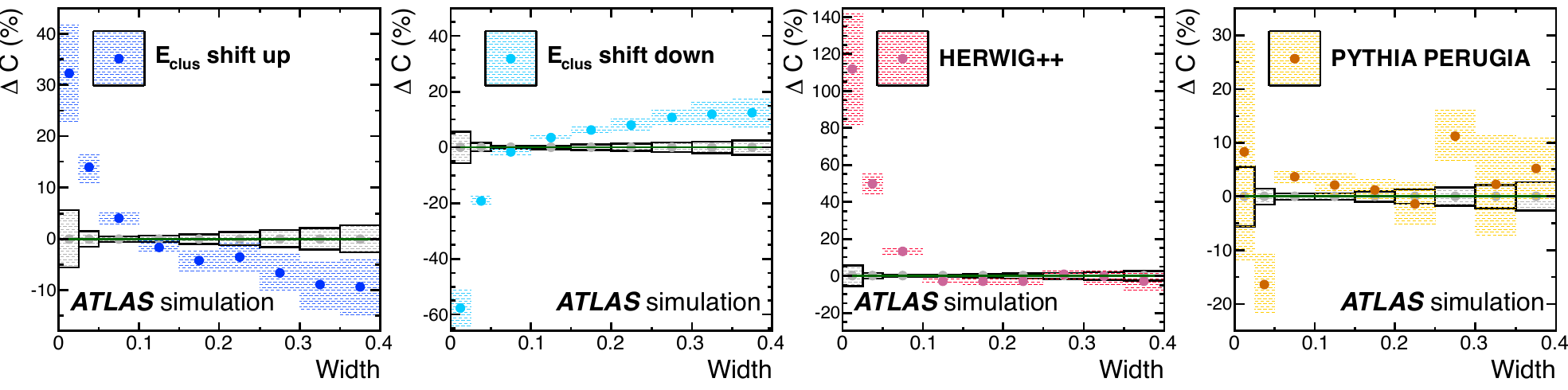}
\includegraphics[scale=0.8,angle=0]{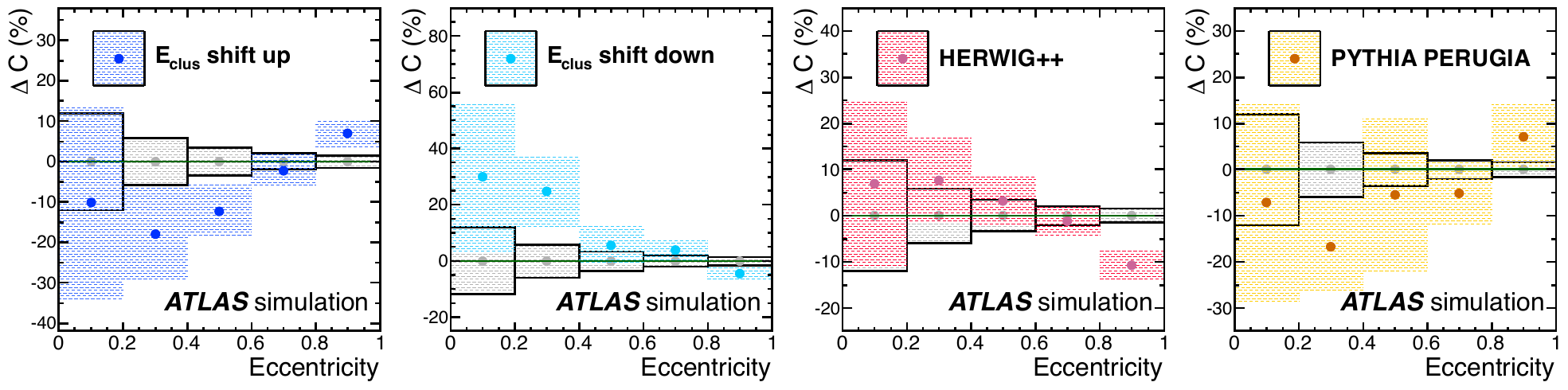}
\includegraphics[scale=0.8,angle=0]{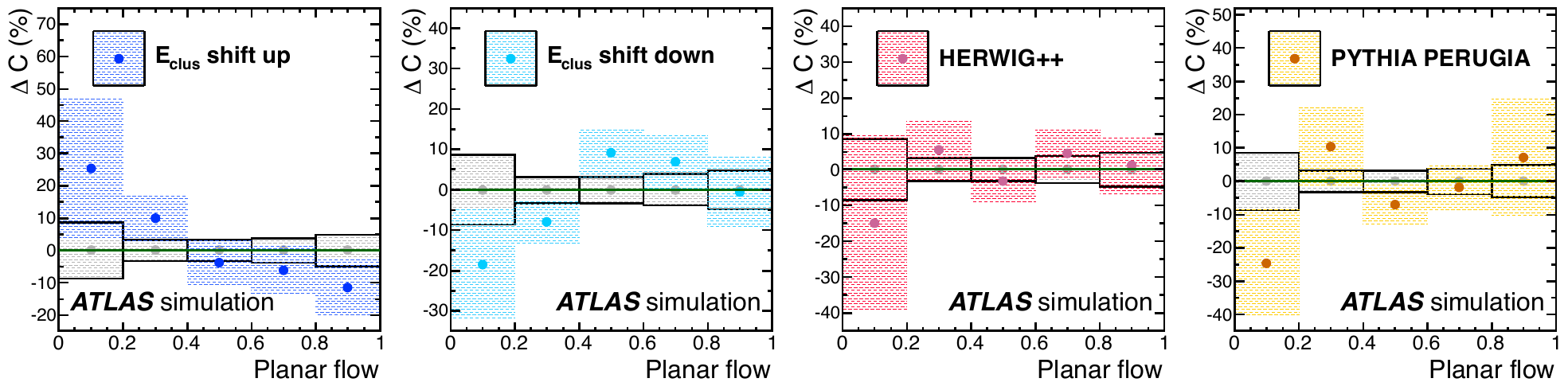}
\end{center}
\caption{The dominant sources of systematic uncertainty on the
  measurements are those resulting in large variations in the detector
  correction factors $C$. These correction factors are found
  bin-by-bin using $R =$ 1.0 jets in a \pythia\ \ambt\ sample with
  upward and downward variations of the cluster energy scale (first
  and second columns), and by using \herwigpp\ (third column) and
  \pythia\ \perugia\ (fourth column) in place of \pythia\ \ambt. The
  differences $\Delta C$ found when comparing the correction factors
  obtained with the baseline \pythia\ \ambt\ sample are shown here for
  each of the properties measured in $R =$ 1.0 jets. The shaded bands indicate the statistical uncertainties.}
\label{fig:sys10det}
\end{figure*}

A primary source
of the uncertainty on the mass measurements is due to
the observed differences in the detector correction factors between
\herwigpp\ and \pythia, with uncertainties ranging between 10--20$\%$ as shown
in \figref{sys6det} and \figref{sys10det}. 

\subsection{Uncertainties on the detector material description and
cluster reconstruction\label{sec:sys_det}}

Performance studies~\cite{ang_sys} have shown that there is
excellent agreement between the measured positions of clusters and tracks in data, indicating no
systematic misalignment between the calorimeter and inner detector.
 The Monte Carlo modeling of the position of clusters with respect to
 tracks is also good, indicating that the detector simulation models
 the calorimeter position resolution adequately; 
however, there remains a small discrepancy between data and Monte Carlo in the
mean and RMS of the track-cluster separation. This source of
uncertainty is taken into account by (Gaussian)
smearing the positions of simulated clusters in $\eta$ and $\phi$ by 5 mrad. This smearing is done independently in $\eta$ and $\phi$, and
the impact on the measurement of the correction factors for each
observable, bin-by-bin, is quantified by taking the
difference, $\Delta C_{i}$, between the correction factors obtained before
and after the position smearing. Smearing the positions in $\eta$ and
$\phi$ results in small $\Delta C_{i}$ for mass and shapes alike,
introducing uncertainties that do not exceed 5$\%$ in any bin.

The variation on the CES follows the procedure used by previous studies~\cite{shapes} according to:

\begin{equation}\label{eq:ces}
p_{\mathrm{clus, new}}=p_{\mathrm{clus}}\times \left(1\pm0.05\times \left(1 +\frac{1.5}{p_{\mathrm{T}}/\mathrm{GeV}}\right)\right)
\end{equation}
\noindent
where $p_{\mathrm{clus}}$ is each component of the cluster's four-momentum and
\pT\ is the cluster \pT\ in GeV. The CES is varied up and down
independently for each momentum component of each cluster, and the correction factors are recalculated in each case
as before. The CES is a large source of systematic uncertainty in
the measurement of mass (of order 20$\%$ across the mass range) and
width (of order 10$\%$ beyond the first two bins). The effects of
varying the CES are in general smaller for the eccentricity, planar
flow and angularity measurements, all of which are made on high mass
jets only. The effects of varying the CES on all observables are shown
with the label $E_{\mathrm{clus}}$ in \figref{sys6det} for $R =$ 0.6 jets and
\figref{sys10det} for $R =$ 1.0 jets. 

The uncertainty introduced as a
result of losing energy due to dead material in the
detector is taken into account by discarding a fraction of low energy ($E <$~2.5~GeV) clusters, following the technique and utilizing the
observations of a previous study of the single hadron response
at $\sqrt{s} = 900$~GeV~\cite{zach_hadrons}. Clusters are not included
in jet reconstruction if they satisfy:

\begin{equation}\label{eq:drop_clusters}
r \;\leq \; \mathcal{P}(E_{0}) \times e^{-2E}, 
\end{equation}

\noindent
where $r$ is a random number $r \in (0,1]$, $\mathcal{P}(E_{0})$ is the
measured uncertainty ($28\%$) on the probability that a particle does
not leave a cluster in the calorimeter, and $E$ is the cluster energy in GeV. 
The impact on the measurement of each observable is quantified by
comparing the correction factors before and after this dropping of low
energy clusters. The impact of this variation is small, resulting in a
contribution to the systematic uncertainty less than a few
percent in all measurements.

\subsection{Uncertainties on the pileup corrections\label{sec:sys_pu}}

There is a statistical uncertainty on the fit $f(x, p_{\rm{T}},
M)$ describing the pileup correction $\Delta x$ for observable
$x$ in $R =$ 0.6 jets. Dedicated studies have shown that the parameterizations of the pileup
corrections in data and in \pythia\ \ambt\ Monte Carlo with simulated
pileup agree, within the statistical uncertainties, for jets across the
\pT\ range considered. The statistical uncertainties on these fits are
accounted for by implementing
$+1\sigma$ and $-1\sigma$ variations independently,
as shown for mass in \figref{mass_ue_mpi}. The correction factors are
recalculated, and in each case the difference is taken as a
contribution to the systematic uncertainty on the measurement.
This is a small contribution to the overall systematic uncertainty
on the measurements, contributing at most a few percent in bins that are
statistically limited, and is a negligible ($<$~1$\%$) effect elsewhere.

For $R =$ 1.0 jets, the correction factors are scaled using the
  phenomenological predictions described in \secref{pileupscaling}. These
  scaling factors are also calculated in data and in \pythia\ \ambt\ Monte Carlo with simulated
pileup; good agreement is observed, indicating that the effect of pileup
on jets is well-modeled. In the
  case of mass and width, where there is a phenomenological prediction for the scaling,
  this prediction is used for the determination of the nominal scaling factors and the variation is
  taken from the scaling factors found in data. In the case of eccentricity there is no phenomenological prediction for the
  scaling of the pileup corrections with $R$, so the behavior
  observed in data is used. The $R$-scaling of the pileup corrections
  for eccentricity is dependent on jet mass, so the variations found in data across the mass range are taken as
  the systematic variations.

The uncertainties introduced by the pileup corrections
contribute a small amount (in general 1--2$\%$) to the total systematic uncertainties on the
mass, width and eccentricity.

The sources of systematic uncertainty described above are
added in quadrature with the statistical uncertainty in each bin and
symmetrized where appropriate (the contributions from the cluster
energy scale and parameterization of the pileup corrections are
determined separately for upward and downward fluctuations, and so are
not symmetrized).

\section{Results\label{sec:results}}

The distributions of jet characteristics presented in this section are corrected for
detector effects and are compared to Monte Carlo predictions at the
particle level. 
In the case of mass and $\tau_{-2}$, comparison is also made between
data and the eikonal approximation~\cite{PhysRevD.79.074012} of NLO
QCD. 
The results shown here are
available in \hepdata~\cite{hepdatapap,hepdatalink} and the analysis and data are available as a \rivet~\cite{rivetpap,rivetlink} routine.

\subsection{Jet mass\label{sec:qcdresults}}

\begin{figure*}
\begin{center}
\includegraphics[scale=0.4,angle=0]{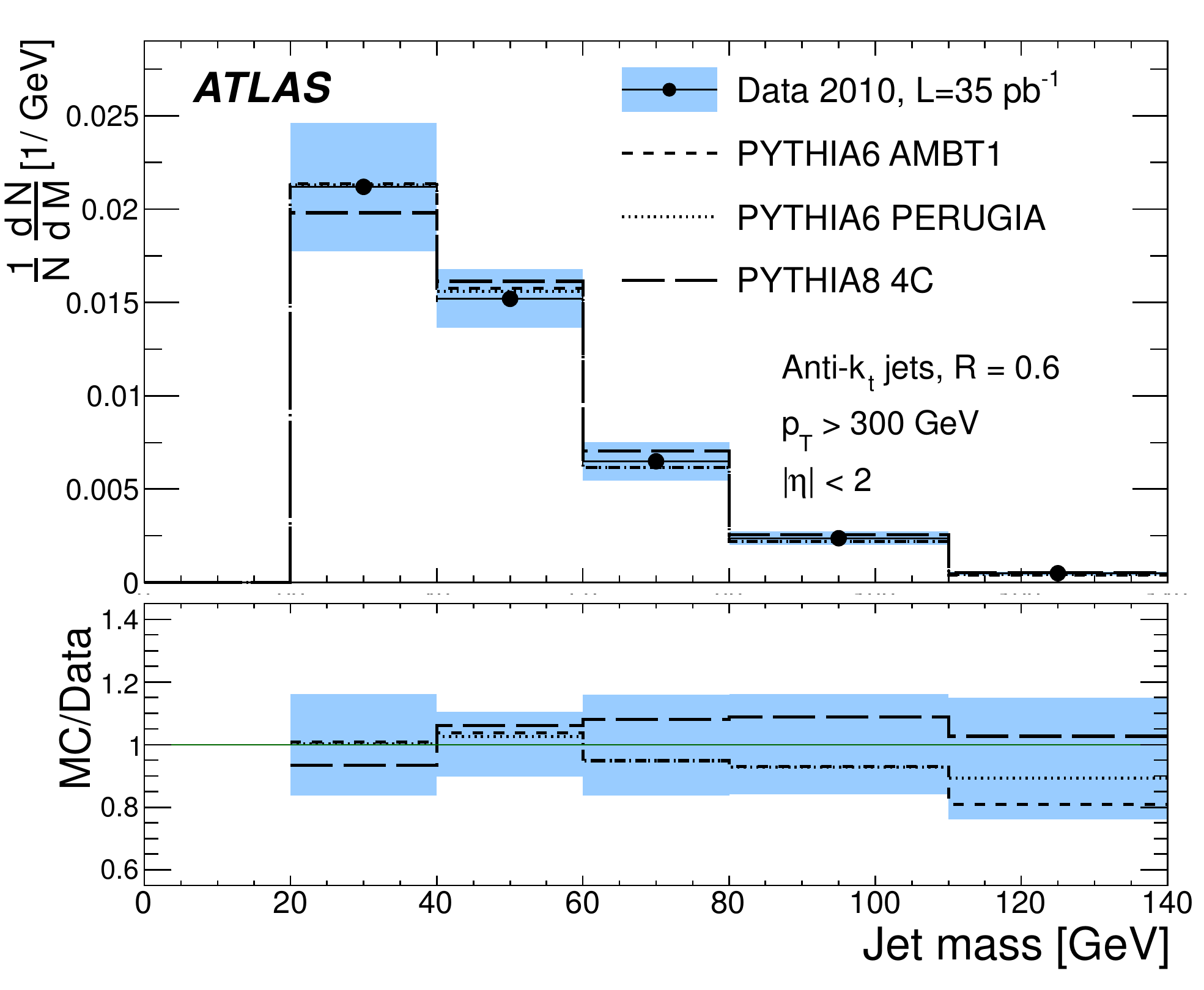}
\hspace{1cm}
\includegraphics[scale=0.4,angle=0]{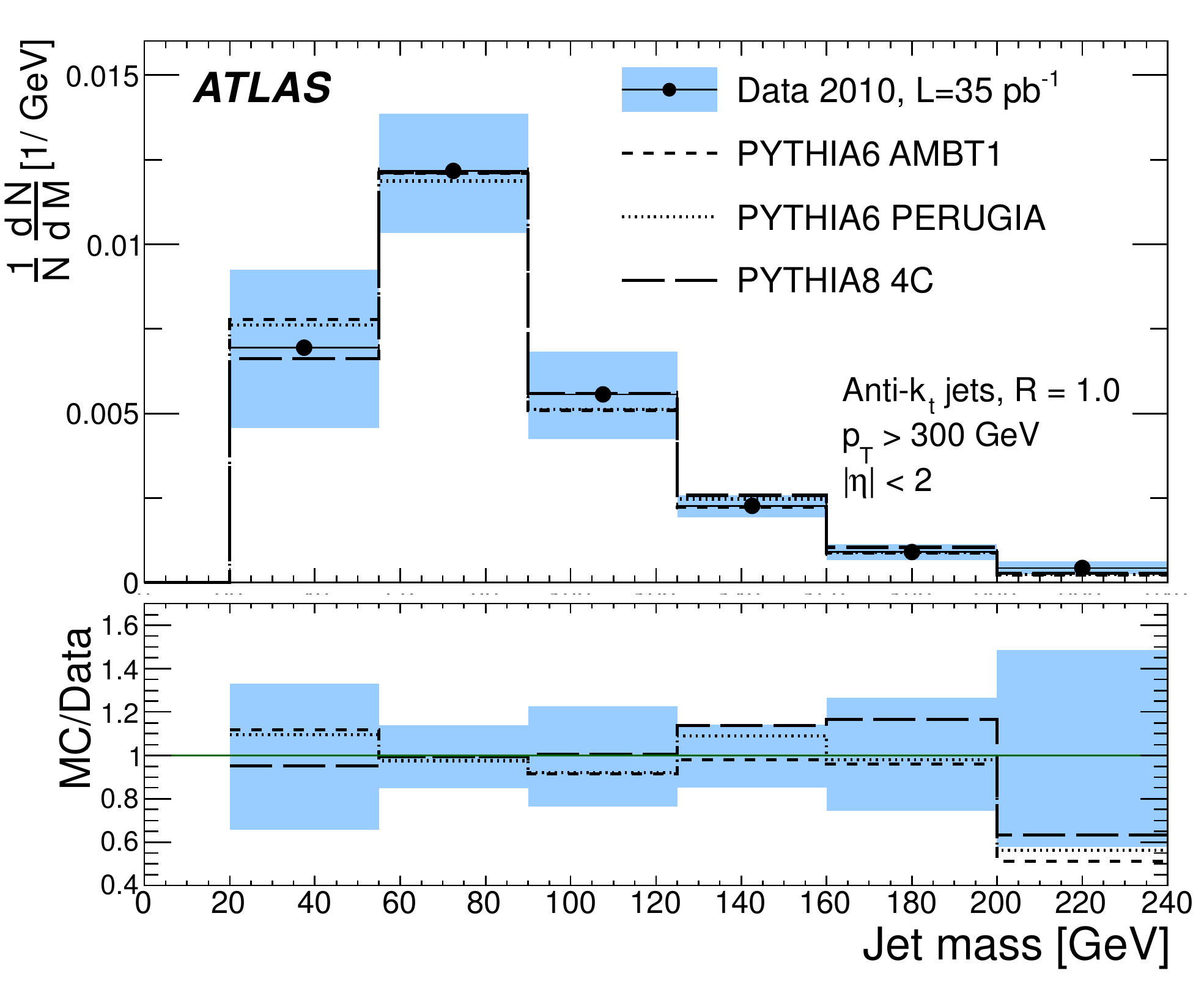}
\includegraphics[scale=0.4,angle=0]{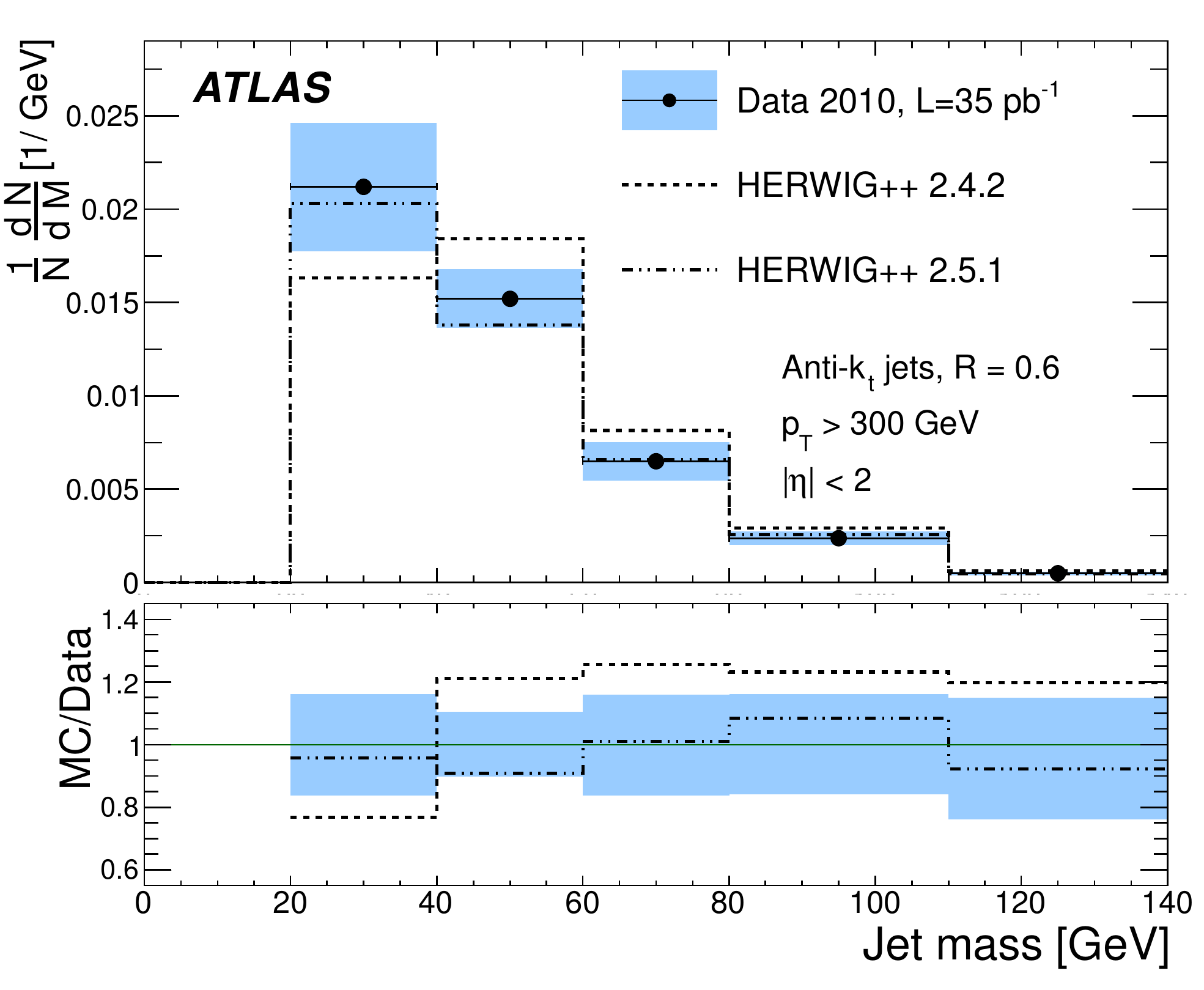}
\hspace{1cm}
\includegraphics[scale=0.4,angle=0]{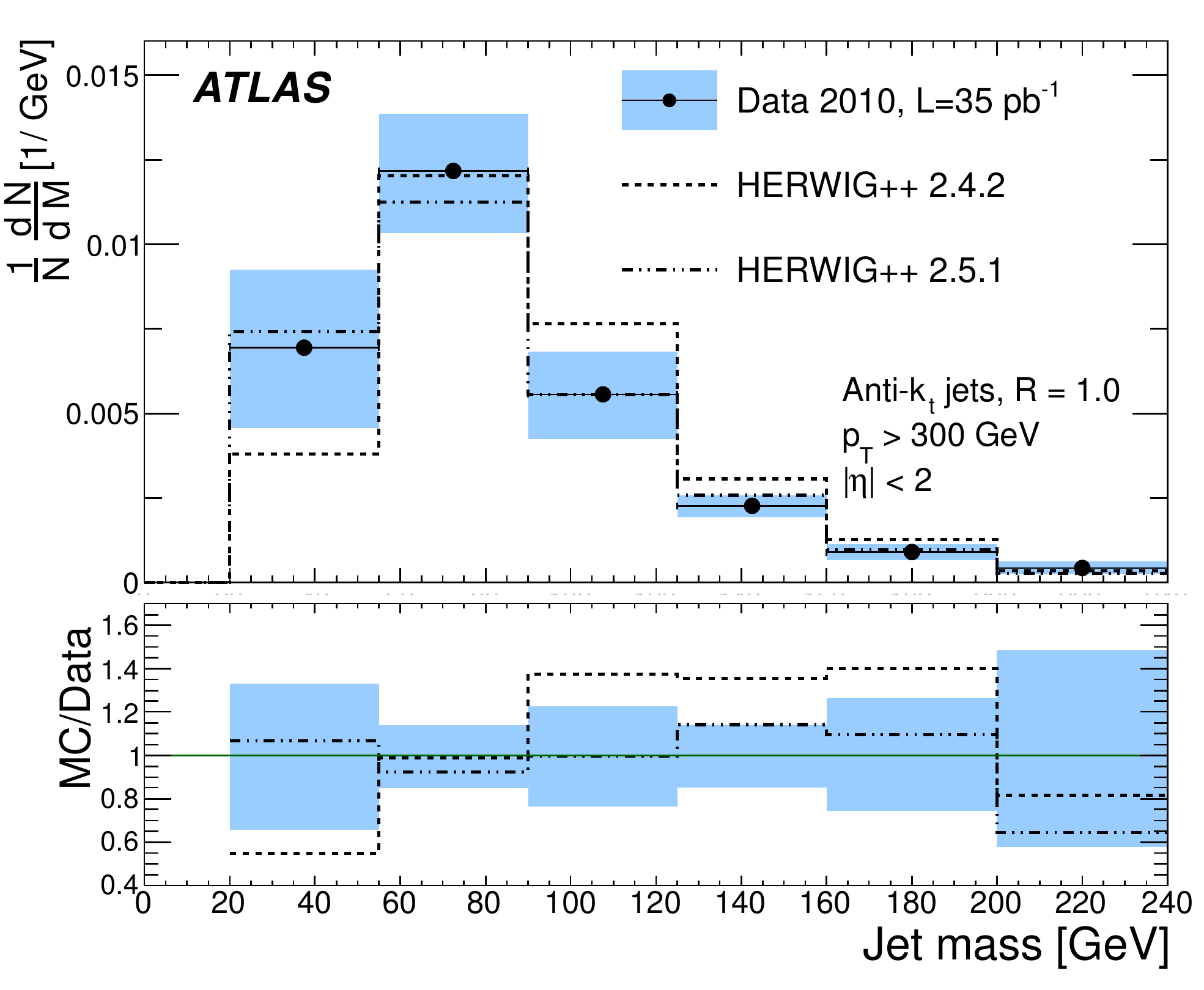}
\includegraphics[scale=0.4,angle=0]{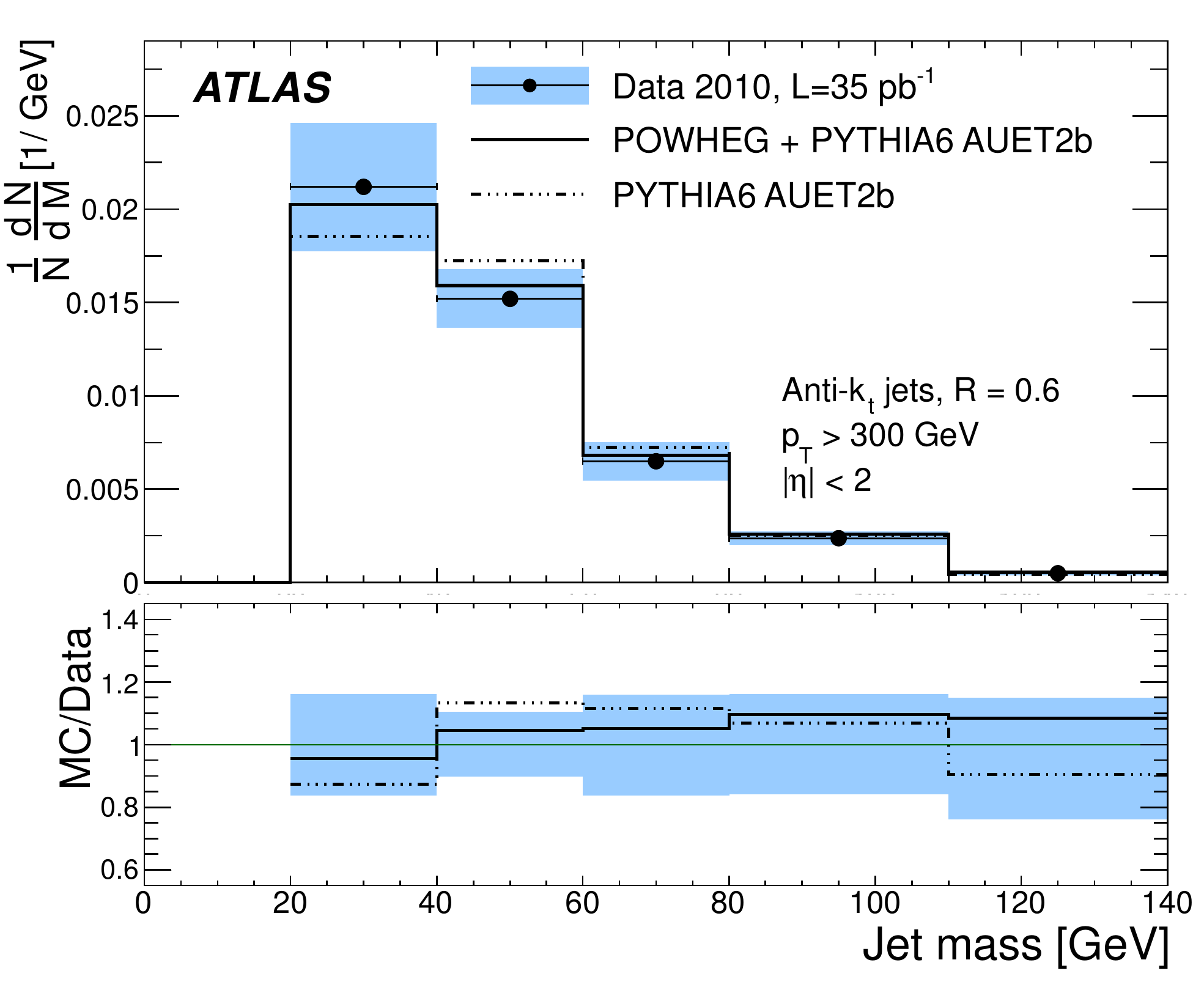}
\hspace{1cm}
\includegraphics[scale=0.4,angle=0]{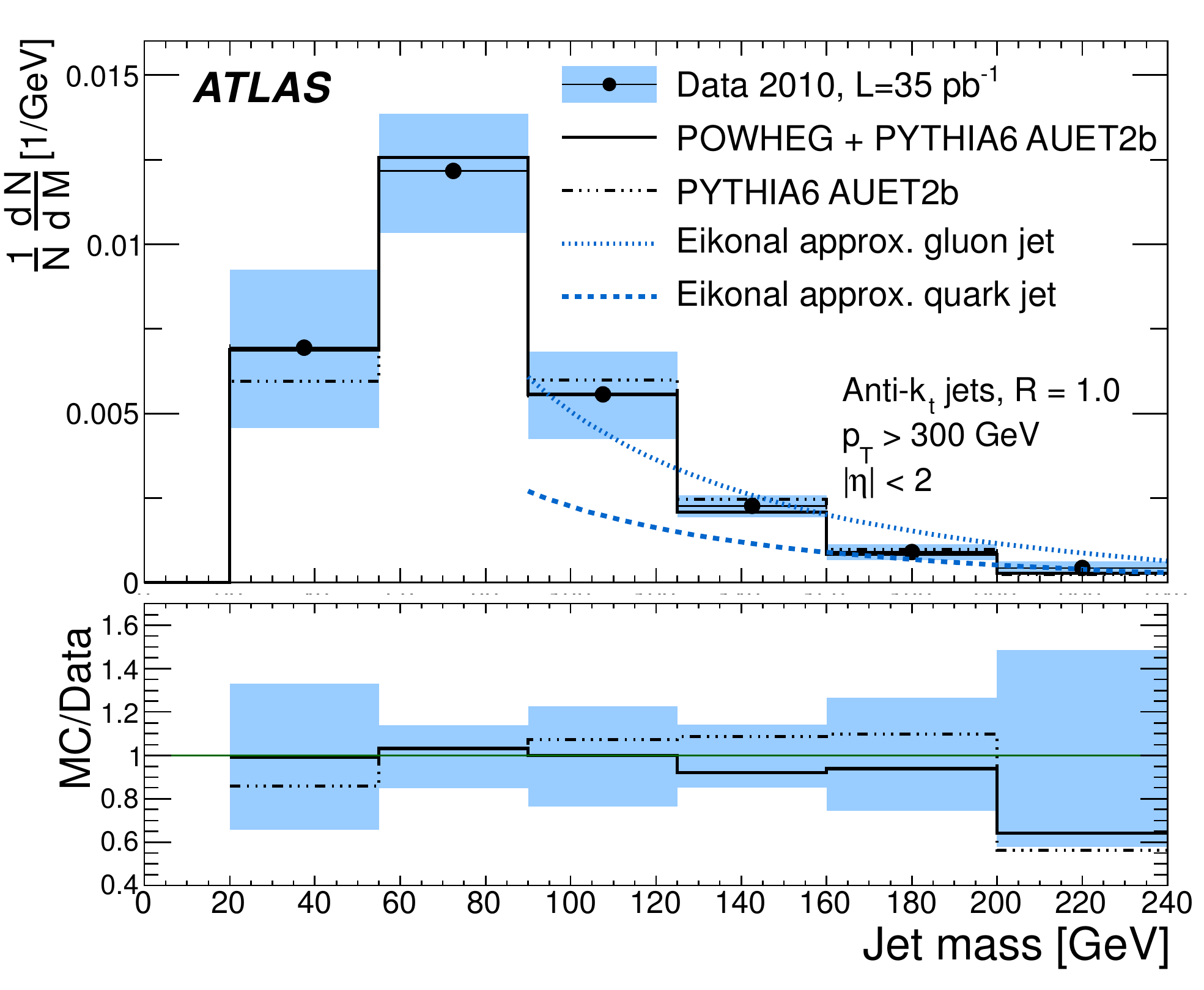}
\end{center}
\caption{The jet mass distributions for leading \pT, \antikt\ $R =$ 0.6 (left) and $R =$ 1.0
  (right) jets in the full 2010 data set, corrected for pileup
  and corrected to particle level. The data are compared to various
  tunes of  \pythia\ 6 and \pythia\ 8 (top), \herwigpp\ 2.4.2 and
  2.5.1 (center) and \pythia\ \aue\ with and without \powheg\ (bottom).  The eikonal approximation of NLO QCD
  for quark and gluon jets is also included for the $R =$ 1.0 case (right, bottom). The shaded bands indicate the sum of
  statistical and systematic uncertainties.
}
\label{fig:finalM}
\end{figure*}

The jet mass distributions are shown in
\figref{finalM} for jets satisfying \pT\ $>$ 300~GeV and $|\eta| <$ 2, corrected 
to the particle level, and the corresponding numerical values are given
in \tabref{tablemass6} and \tabref{tablemass10}. 

In the case of $R =$ 1.0 jets, the data are compared to the calculations for 
jet masses derived at NLO QCD in the 
eikonal approximation: 
\begin{eqnarray}
J \simeq \alpha_{\mathrm{S}}  \frac{  4\,  C_{\mathrm{c}}}{\pi M} \log\left( \frac{1}{z} \tan\left(\frac{R}{2}\right)   \sqrt{4-z^2} \right),
\end{eqnarray}
where $J$ is the value of the jet mass distribution at $M$, $\alpha_{\mathrm{S}} $ is the strong coupling
constant,  $z= M / p_{\mathrm{T}}$, $\mathrm{c}$ represents the flavor
of the parton which initiated the jet and $C_{\mathrm{c}}=\frac{4}{3}$
($3$) for quarks (gluons).
The strong coupling constant is calculated using the \pythia\
prediction of the average jet \pT\ $\simeq 365$~GeV and has the value of  $\alpha_{\mathrm{S}} =0.0994$. 
Theoretical uncertainties for such predictions are sizable (more than 30$\%$)~\cite{PhysRevD.79.074012} in the region above
the mass peak. The lower mass region $M \lesssim$ 90~GeV is strongly
affected by non-perturbative physics and as such cannot be predicted by such calculations. The size and shape of the high-mass tail is in rough agreement with
the analytical eikonal approximation for NLO QCD for jet masses above 90~GeV, with most of
the data points lying between the predictions for quark-initiated and
gluon-initiated jets. QCD LO calculations predict that the jets in
this sample should be roughly 50$\%$ quark-initiated, with this
fraction increasing as a function of the jet \pt\ cut~\cite{quarkgluon}.

Also included in \figref{finalM} are a number of \pythia, \herwigpp\
and \powheg\ predictions
for the jet mass distributions. Unlike the analytical calculation discussed above, 
the Monte Carlo predictions are meaningful down to the low mass region
due to the inclusion of soft radiation and hadronization.
The \pythia\ calculation describes the data well.
The \herwigpp\ 2.4.2 prediction indicates a significant shift to a higher jet mass
that is inconsistent with the data and the other Monte Carlo
predictions, while the more recent \herwigpp\ 2.5.1 generator is in much better
agreement with the data. \powheg+\pythia\ is in good agreement with data within systematic uncertainties across the whole mass range. 

\subsection{\label{sec:reswidth} Width}

\begin{figure*}
\begin{center}
\includegraphics[scale=0.4,angle=0]{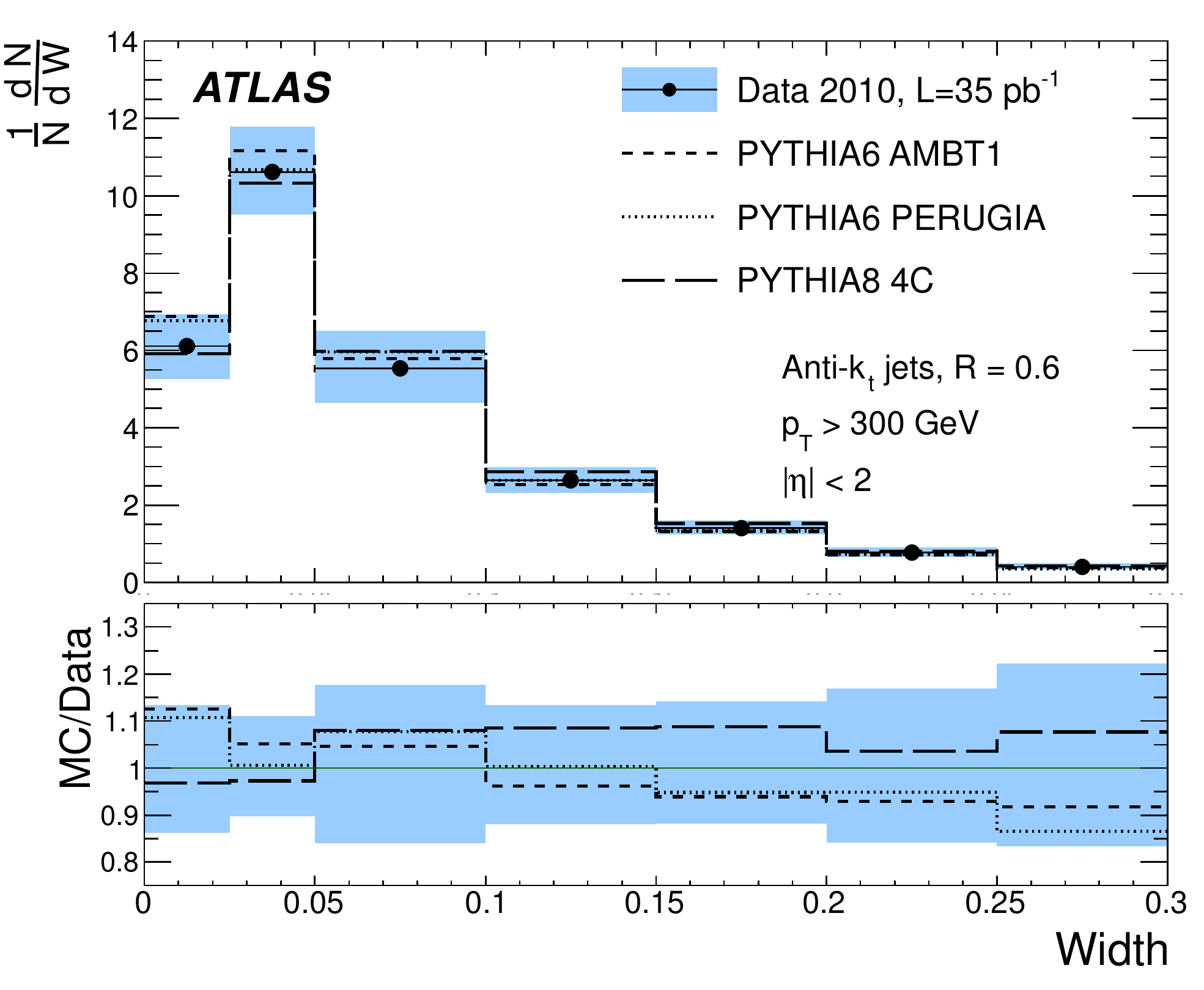}
\hspace{1cm}
\includegraphics[scale=0.4,angle=0]{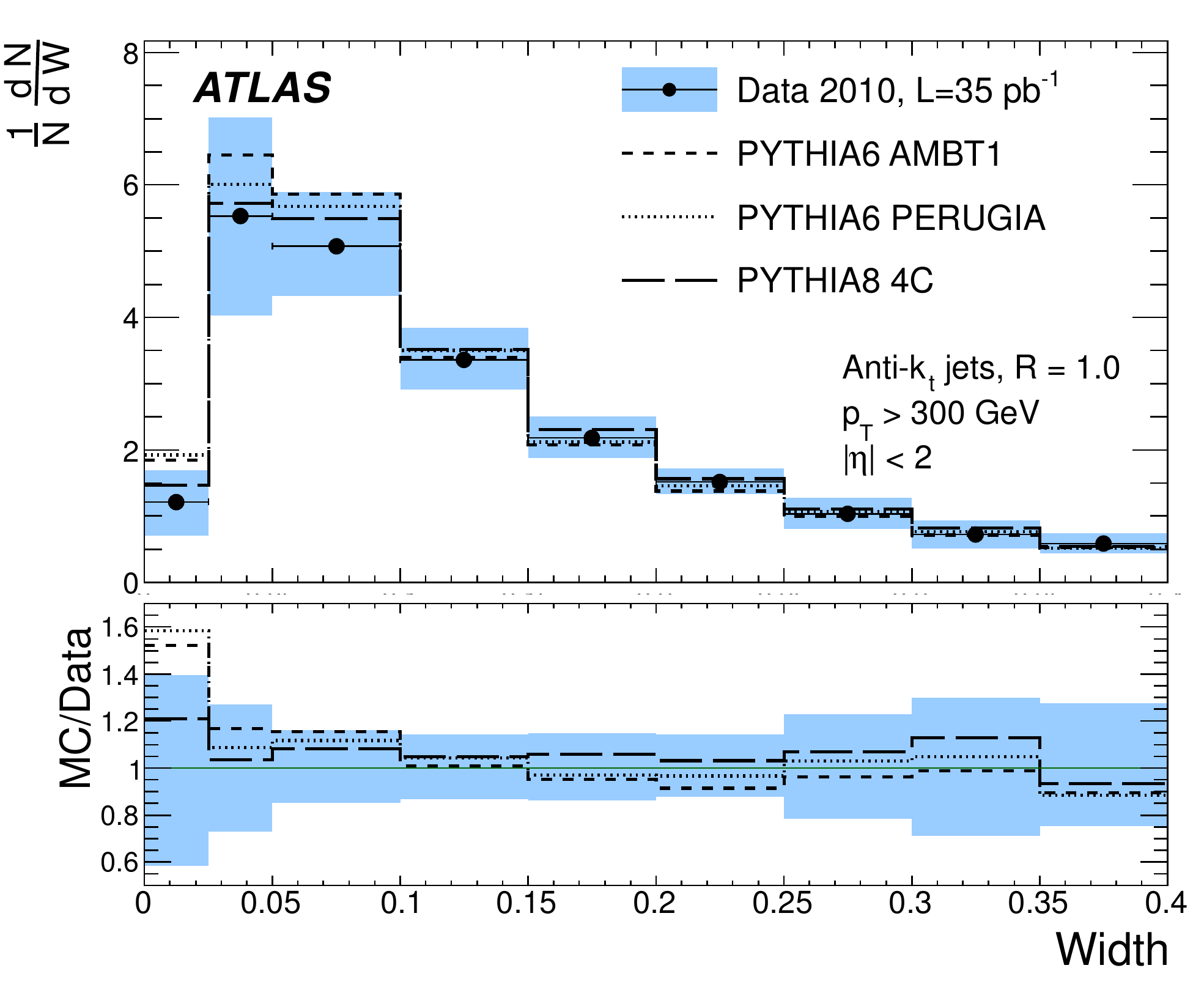}
\includegraphics[scale=0.4,angle=0]{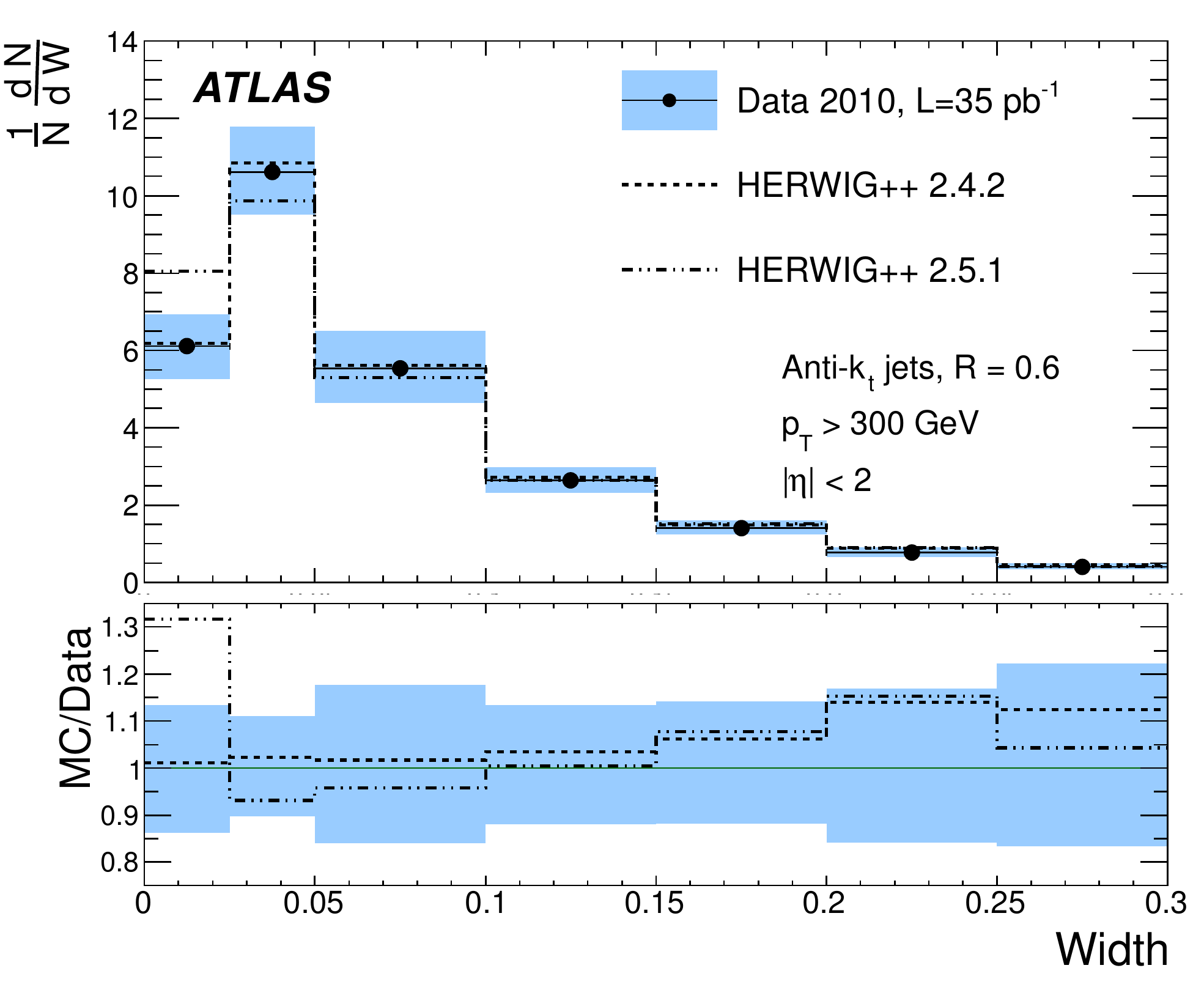}
\hspace{1cm}
\includegraphics[scale=0.4,angle=0]{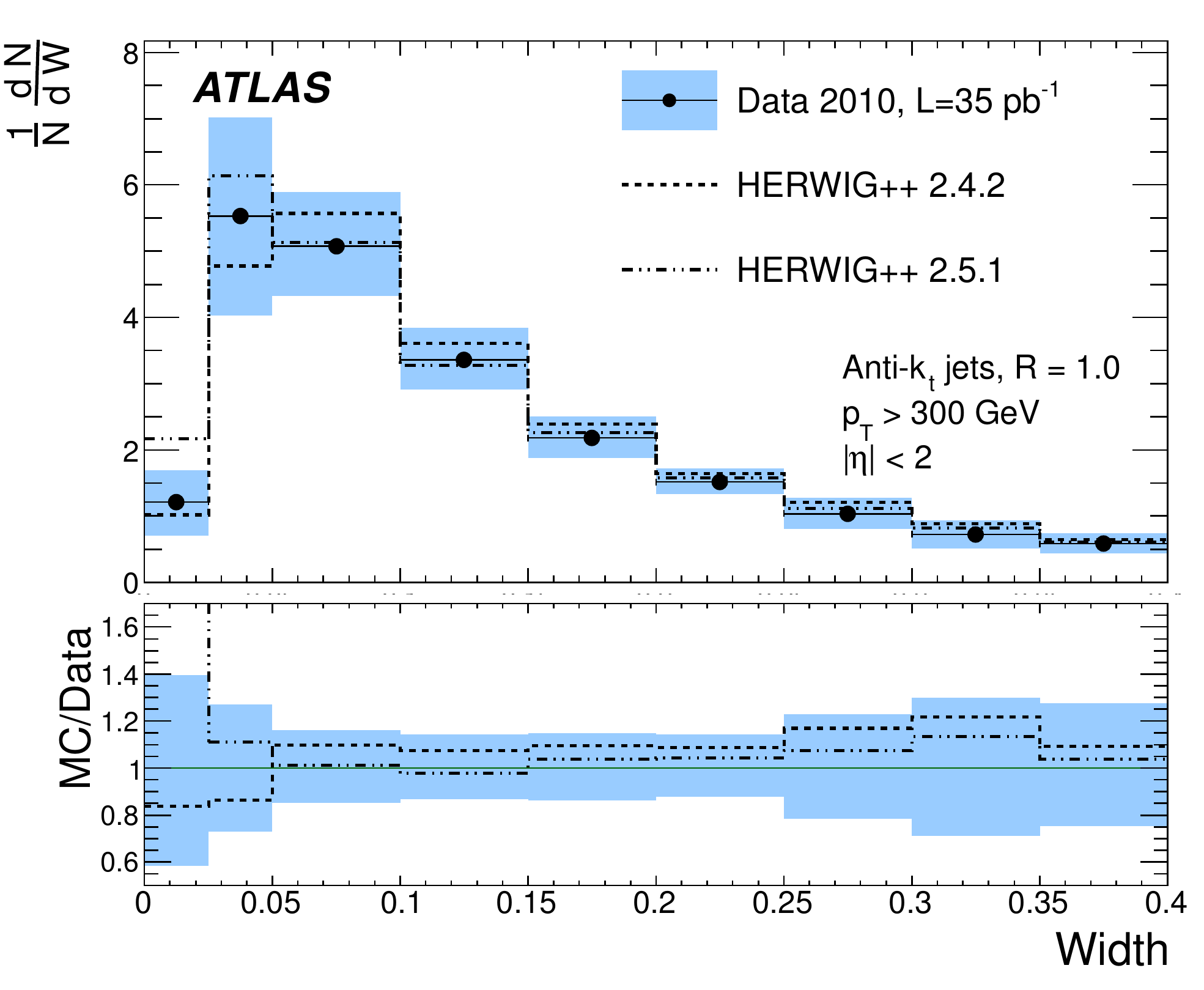}
\includegraphics[scale=0.4,angle=0]{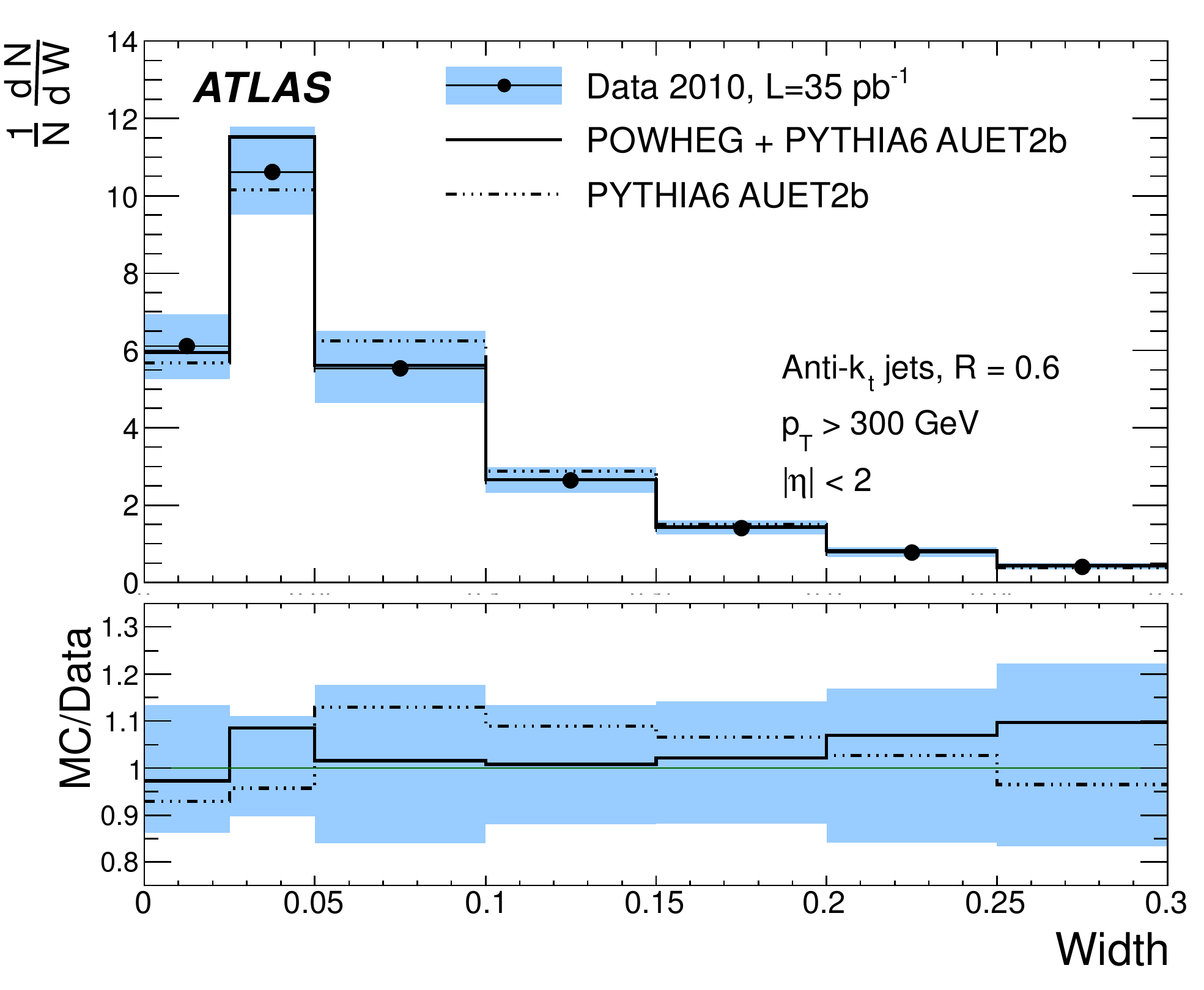}
\hspace{1cm}
\includegraphics[scale=0.4,angle=0]{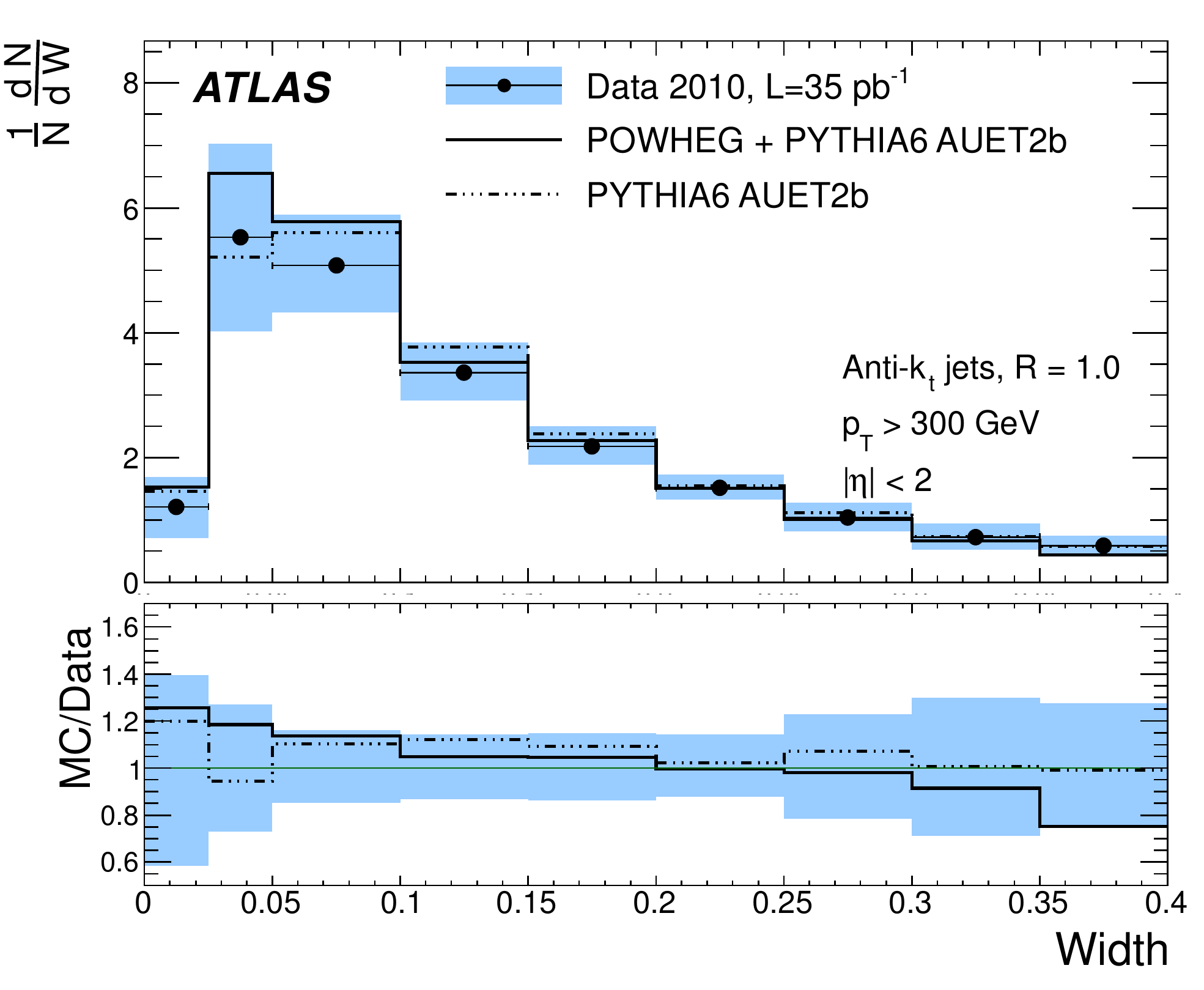}
\end{center}
\caption{The jet width distributions for leading \pT, \antikt\ $R =$ 0.6 (left) and $R =$ 1.0
  (right) jets in the full 2010 data set, corrected for pileup
  and corrected to particle level.
}
\label{fig:finalW}
\end{figure*}

The jet width distributions are shown in
\figref{finalW} for \antikt\ jets reconstructed with distance
parameters of $R =$ 0.6 and 1.0, and the corresponding numerical values are given
in \tabref{tablew6} and \tabref{tablew10}. There is significant
variation between the different Monte Carlo predictions in the first
bin, beyond which there is good agreement
between the distribution measured in data and all the predictions.

\subsection{\label{sec:resecc} Eccentricity}

\begin{figure*}
\begin{center}
\includegraphics[scale=0.4,angle=0]{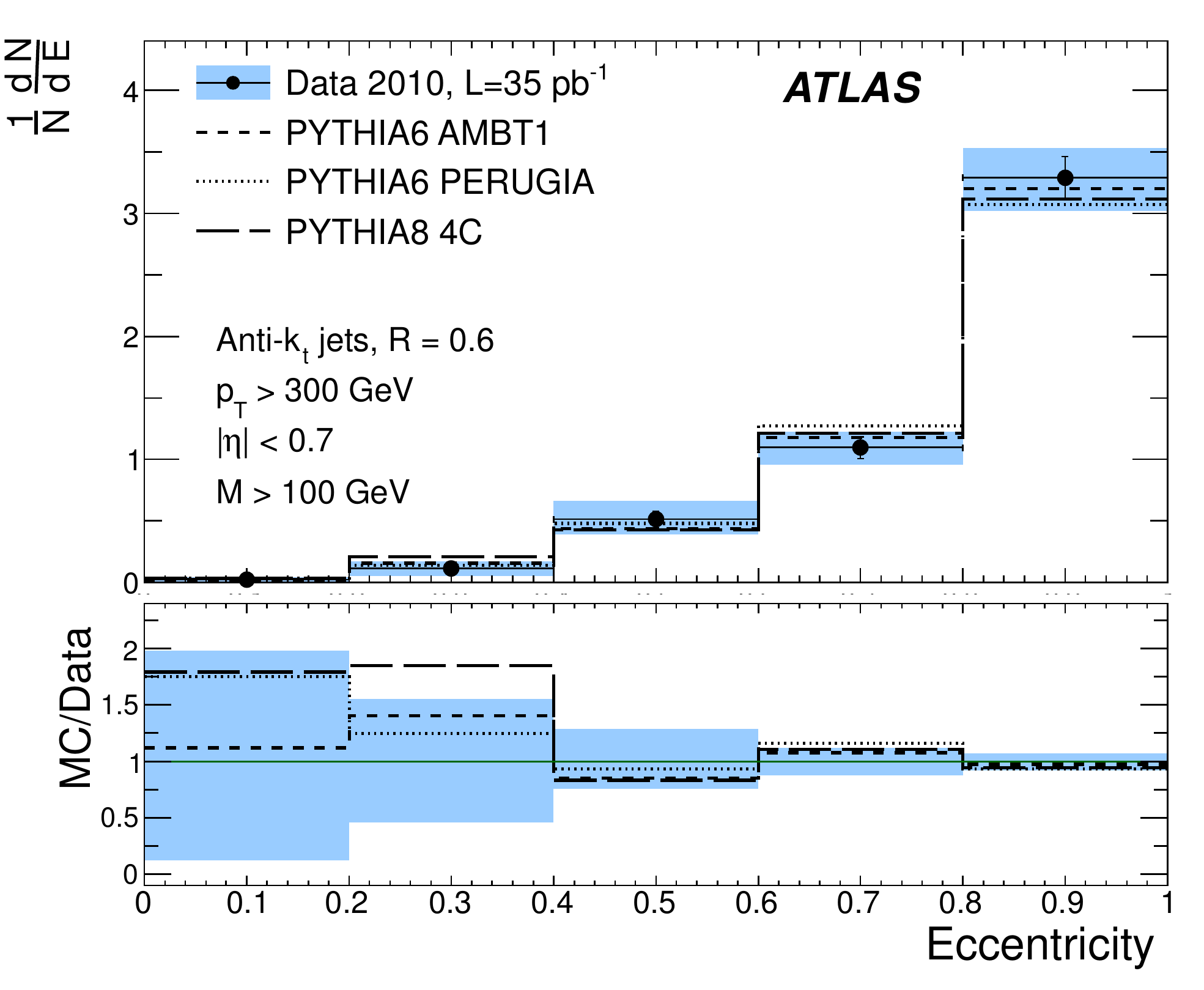}
\hspace{1cm}
\includegraphics[scale=0.4,angle=0]{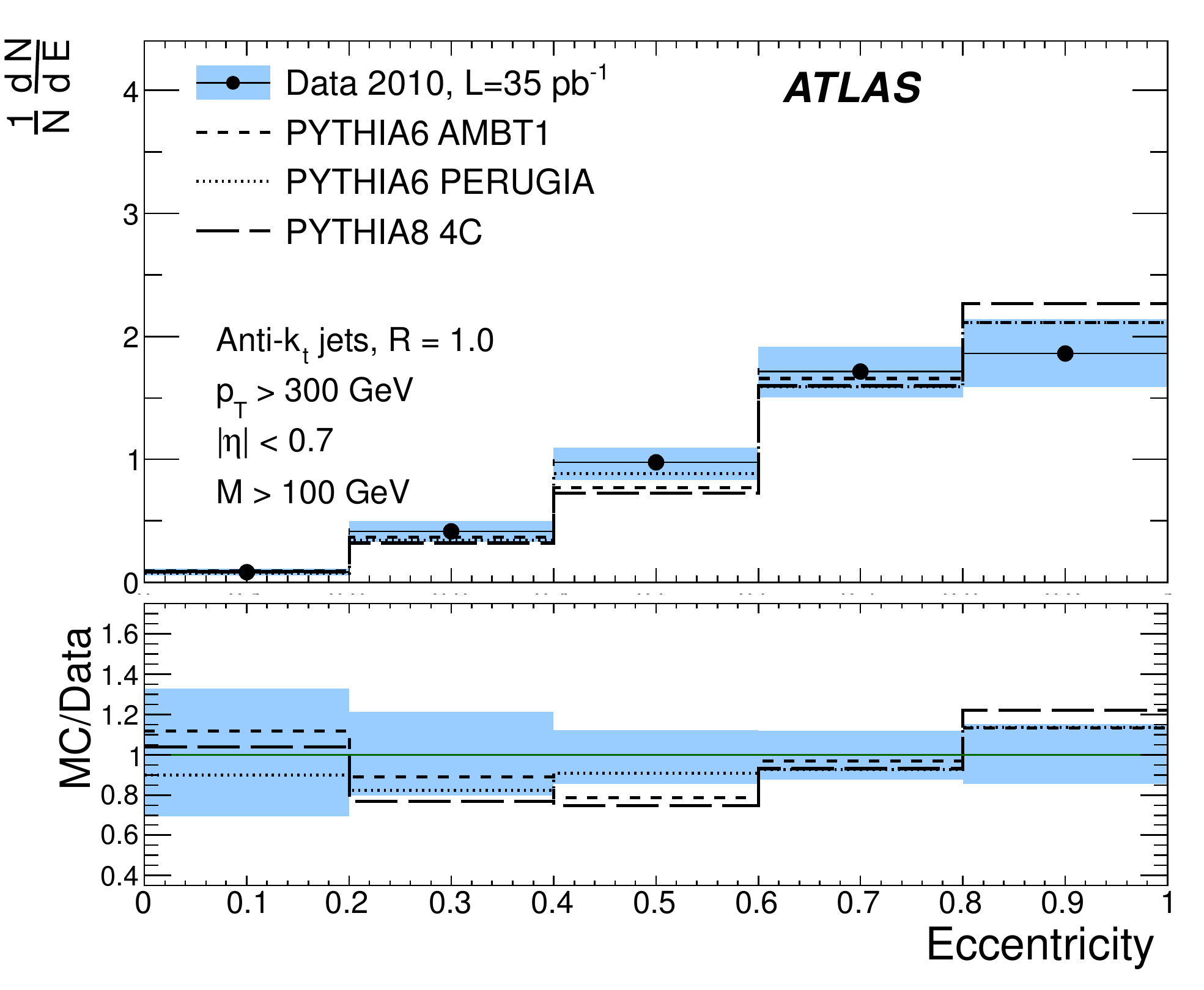}
\includegraphics[scale=0.4,angle=0]{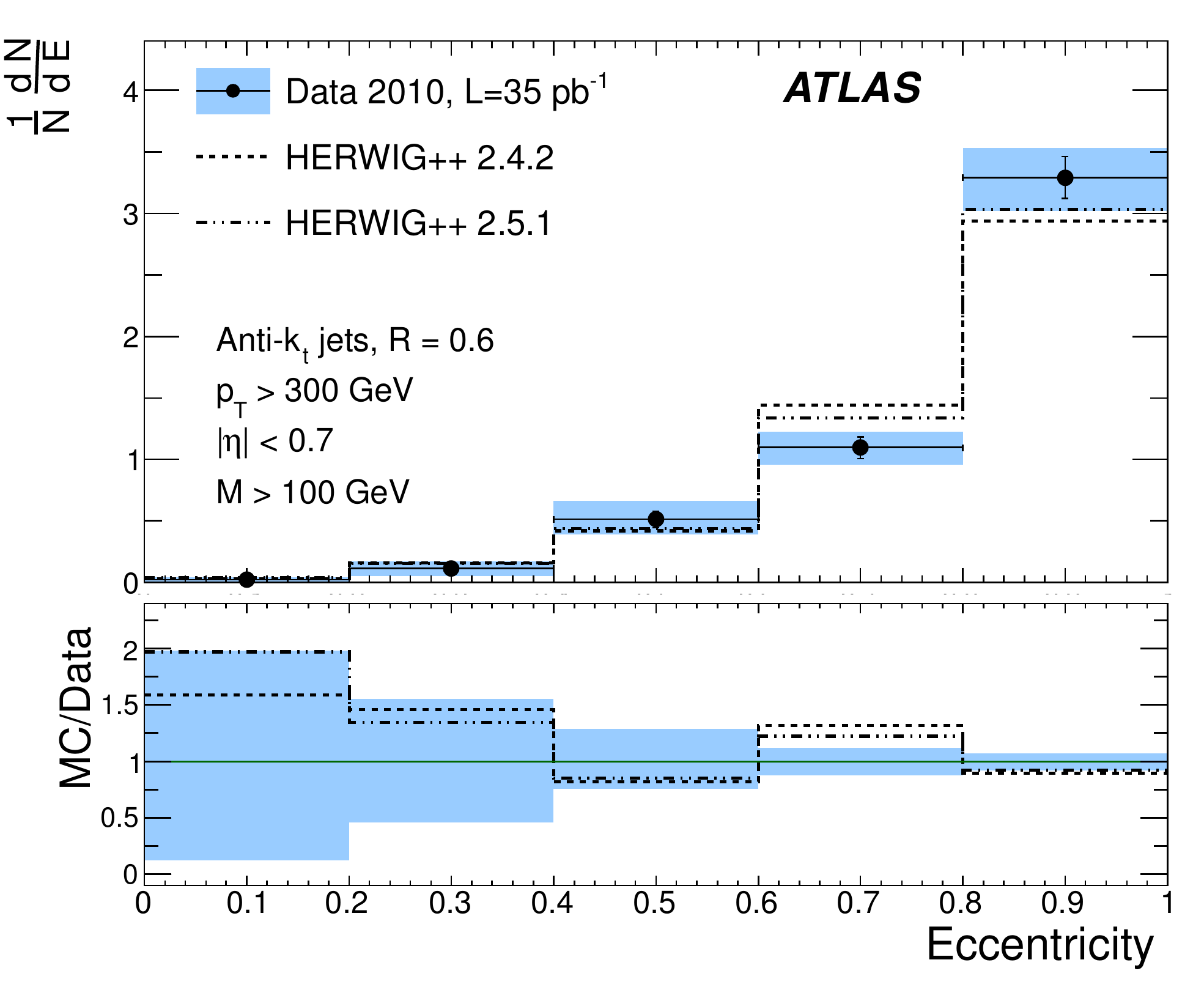}
\hspace{1cm}
\includegraphics[scale=0.4,angle=0]{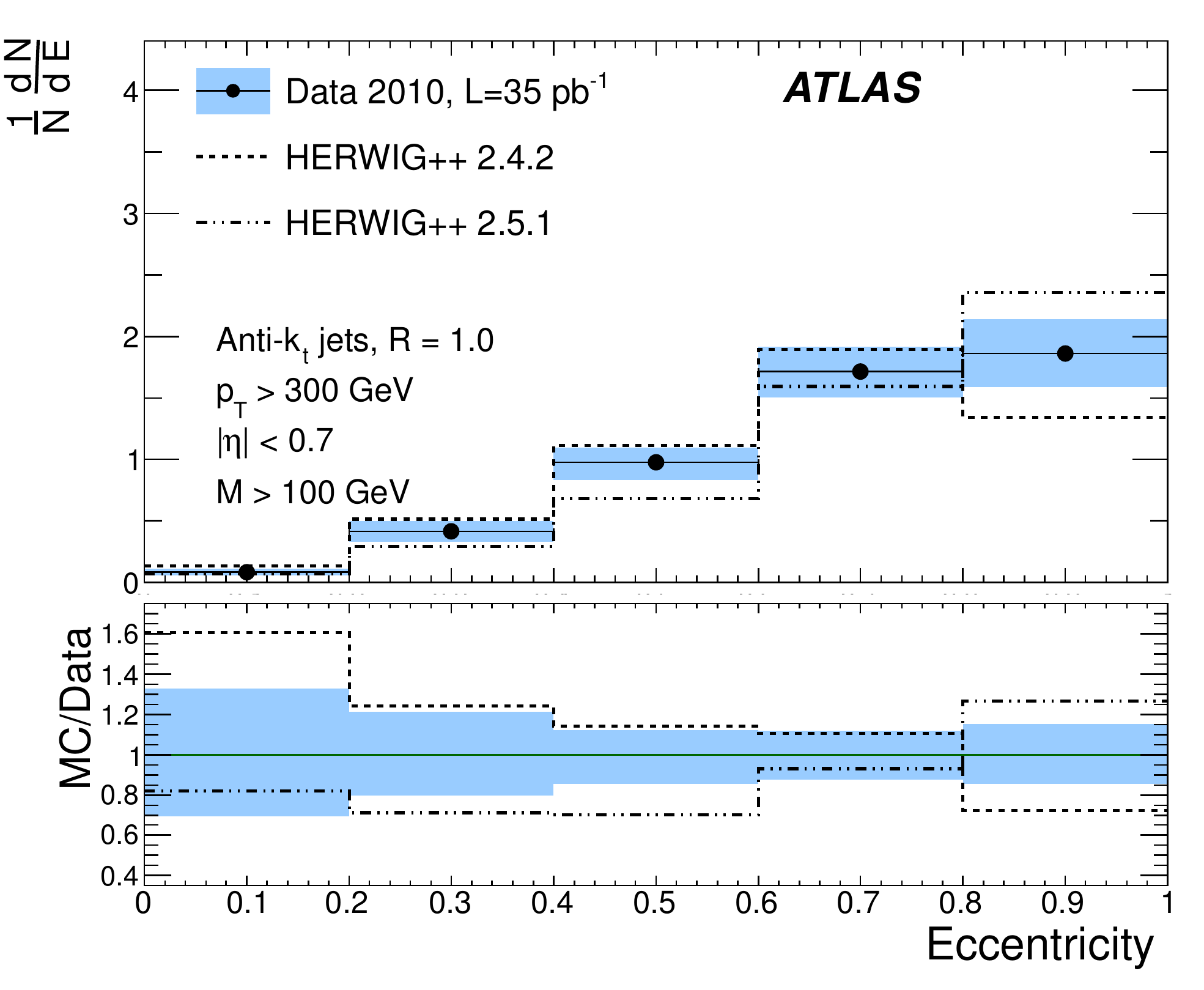}
\includegraphics[scale=0.4,angle=0]{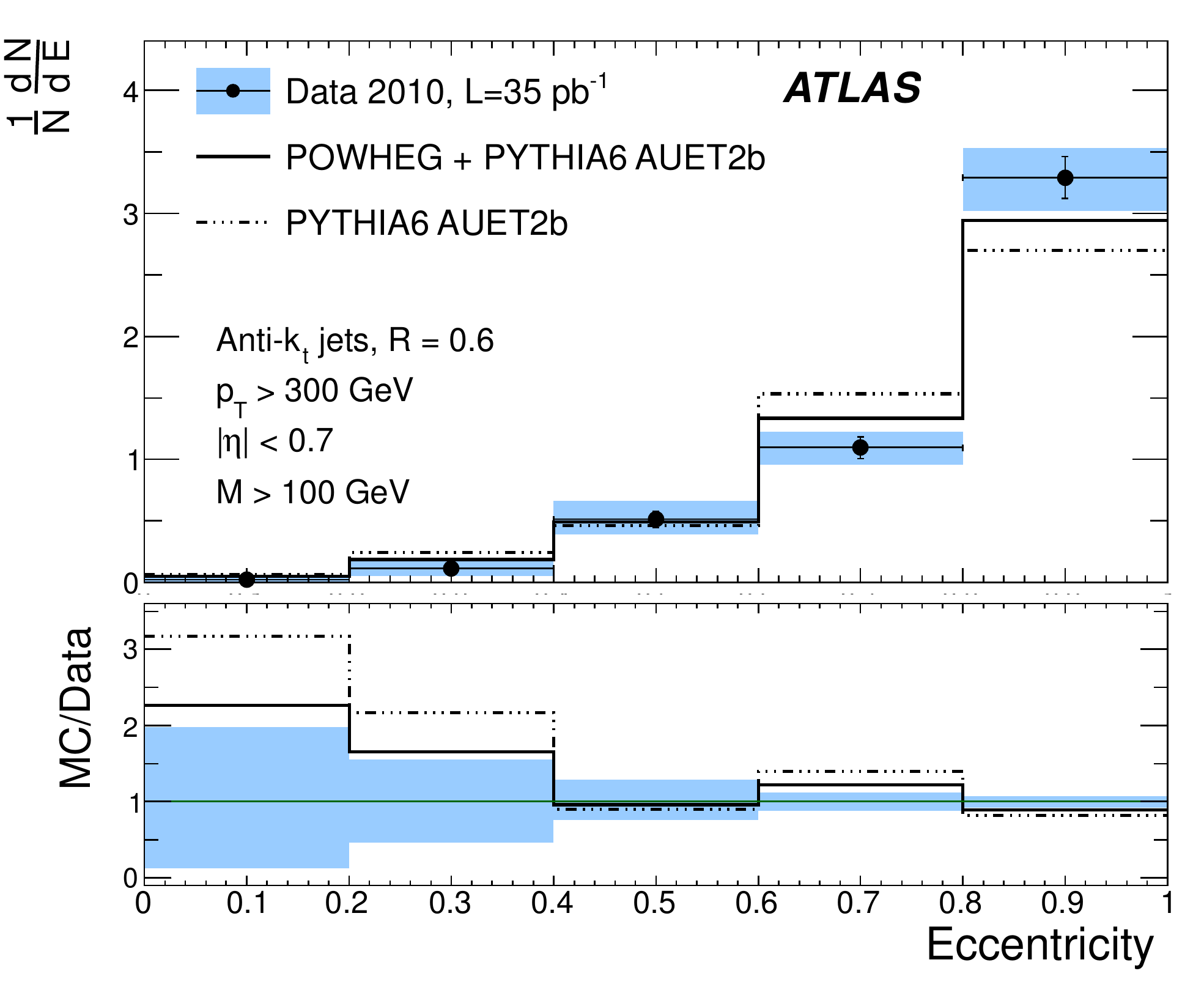}
\hspace{1cm}
\includegraphics[scale=0.4,angle=0]{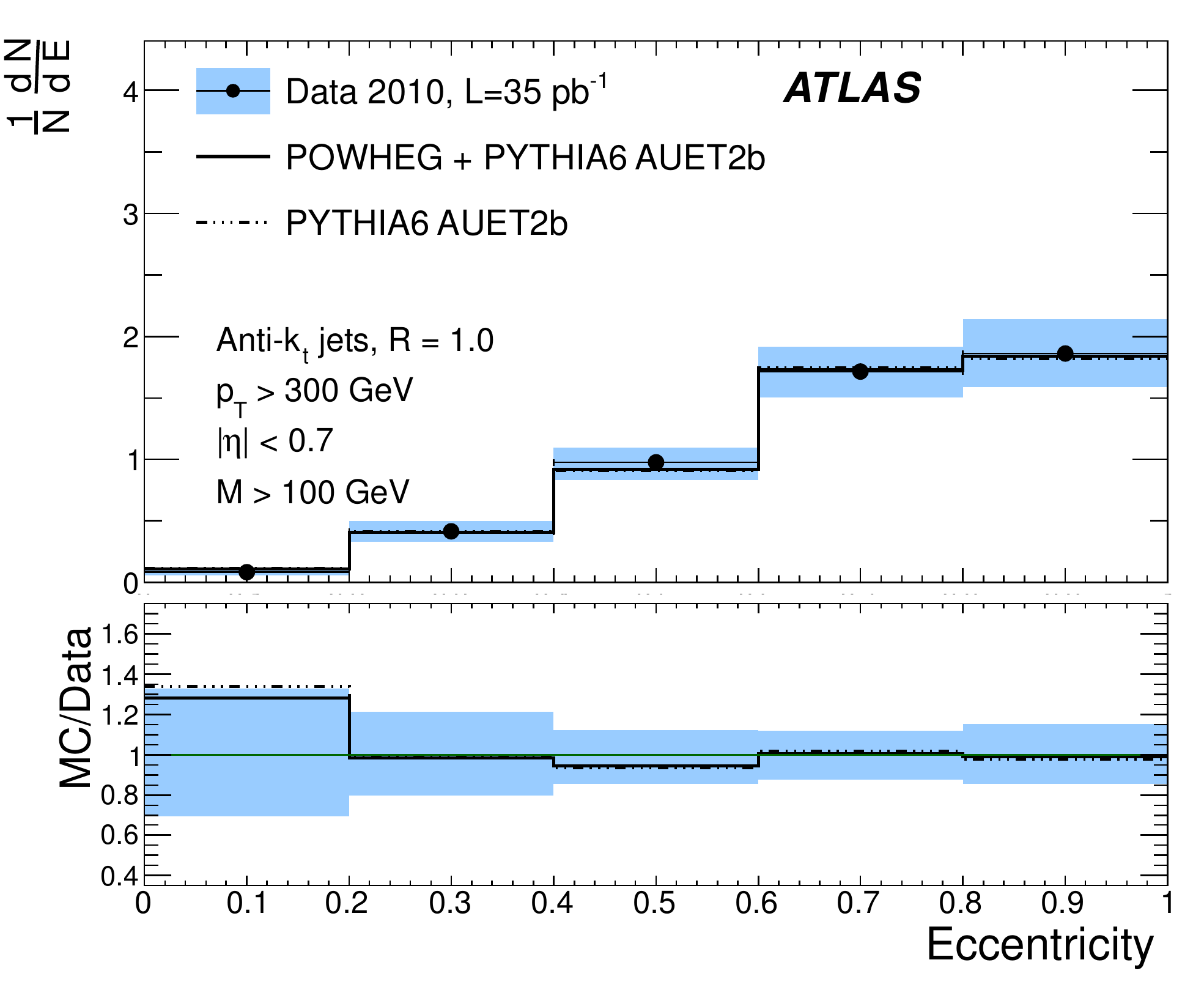}
\end{center}
\caption{The jet eccentricity distributions for high mass ($M >$
  100~GeV), leading \pT,  \antikt\ $R =$ 0.6 (left) and $R =$ 1.0 (right) jets
 in the full 2010 data set, corrected for pileup and corrected to
 particle level.  
}
\label{fig:final_ecc}
\end{figure*}

The eccentricity distributions for high mass ($M >$ 100~GeV) \antikt jets
reconstructed with distance parameters of $R =$ 0.6 and $R =$ 1.0 are shown in
\figref{final_ecc}, and the corresponding numerical values are given
in \tabref{tablee6} and \tabref{tablee10}. The Monte Carlo predictions generally describe
the data, while some small discrepancies can be observed between the various
predictions and between predictions and data.

\subsection{\label{sec:respf} Planar flow}

\begin{figure}
\begin{center}
\includegraphics[scale=0.4, angle=0]{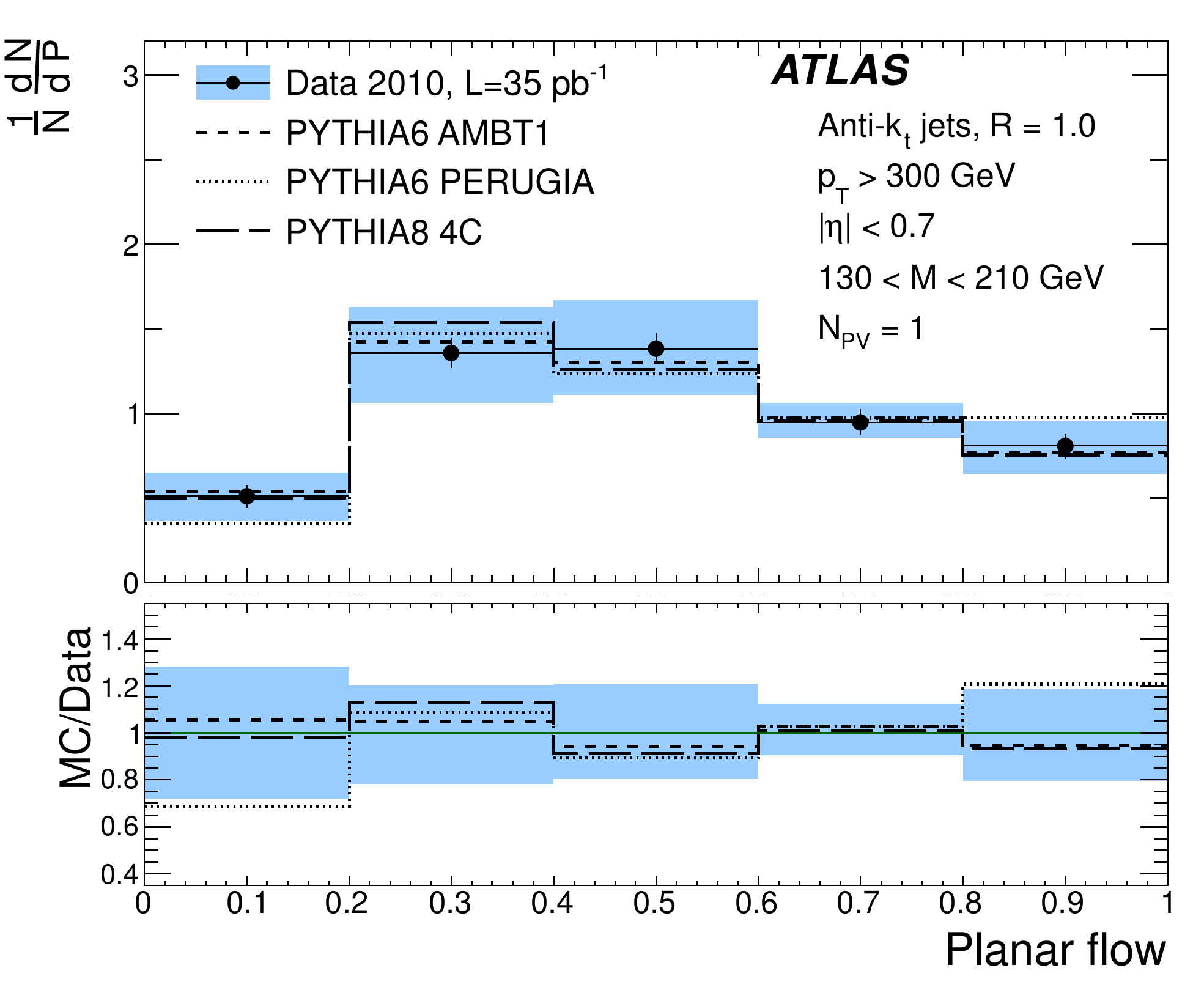}
\includegraphics[scale=0.4,angle=0]{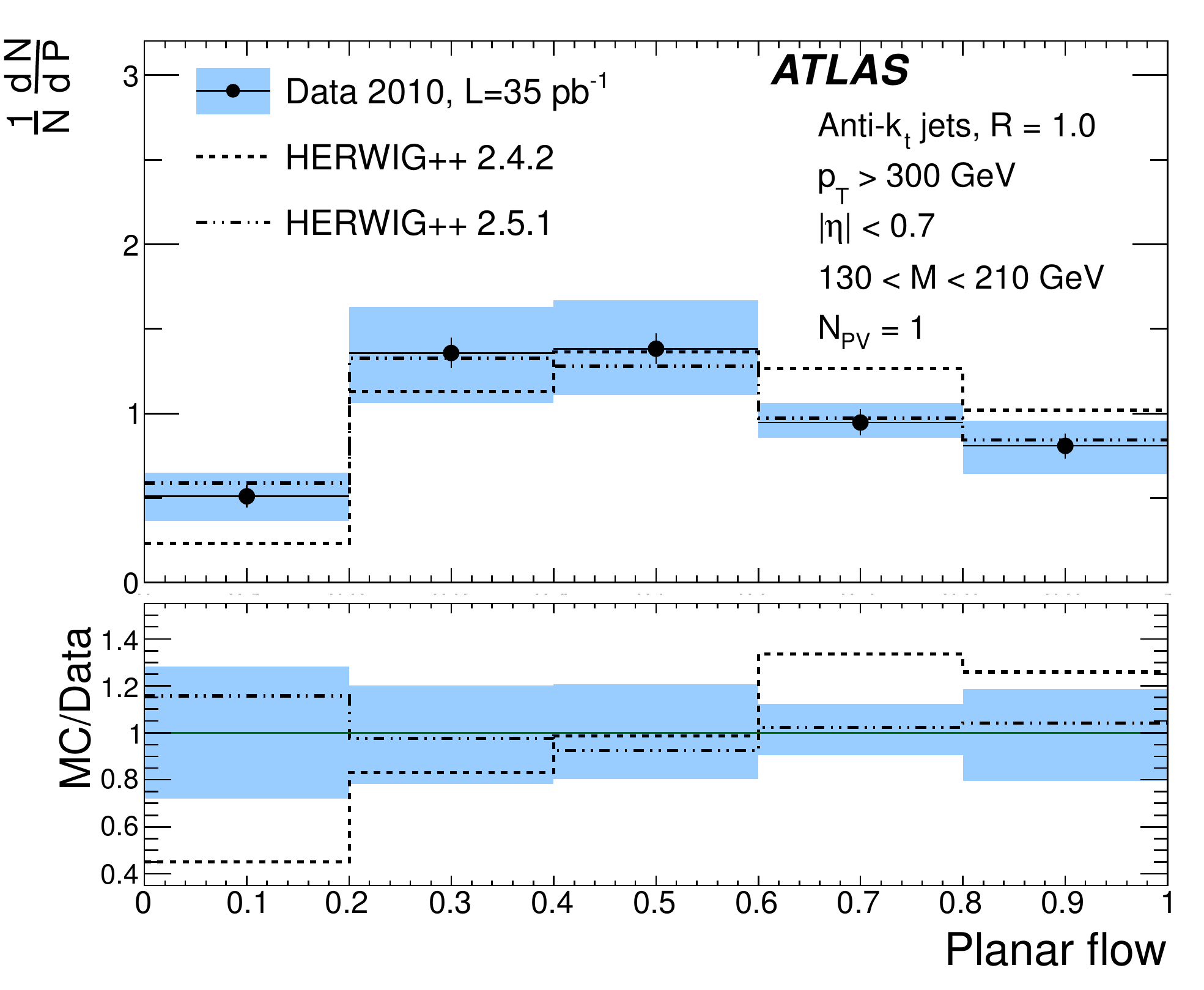}
\includegraphics[scale=0.4, angle=0]{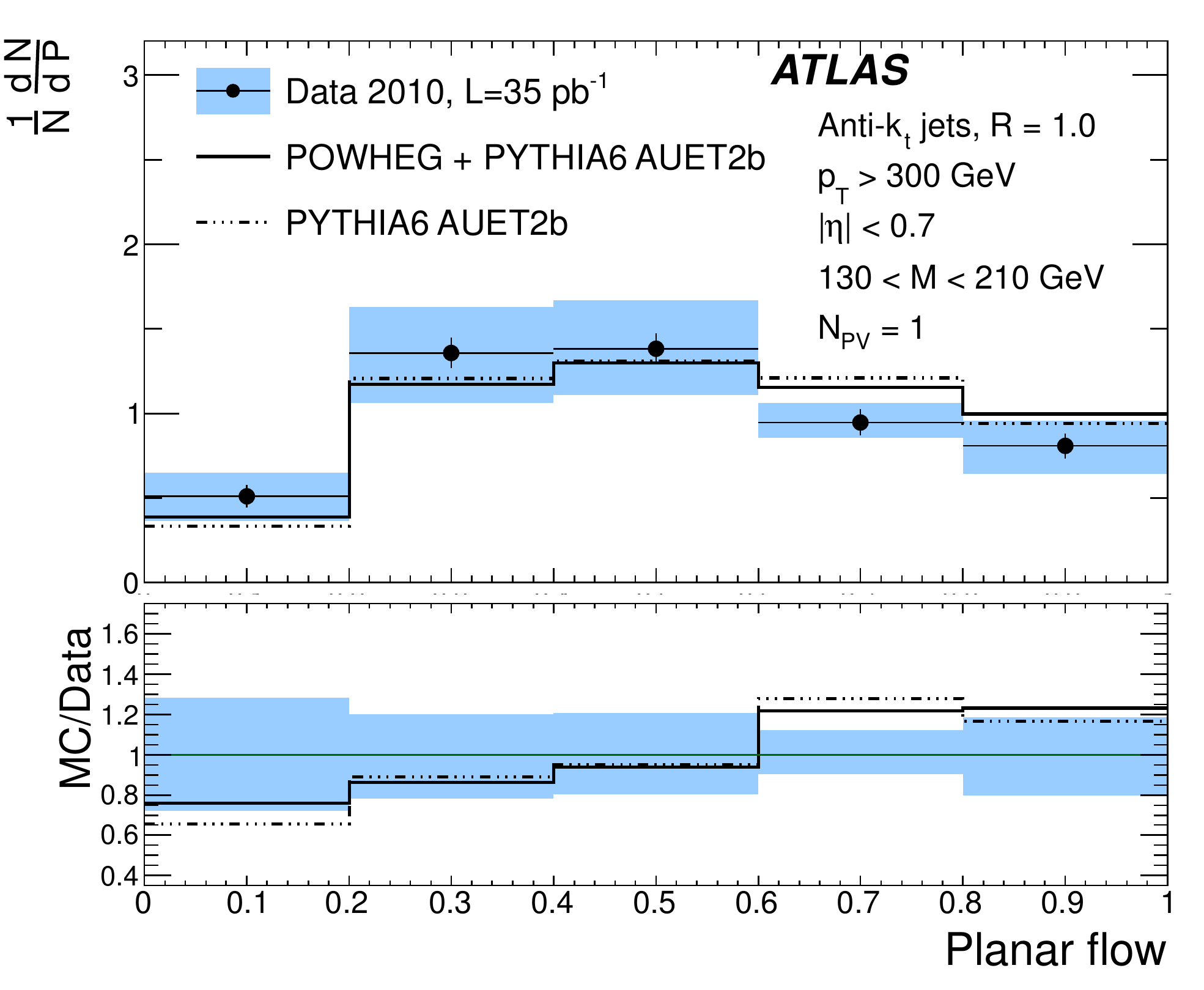}
\end{center}
\caption{The jet planar flow distributions for high mass (130~$< M
  <$~210~GeV), leading \pT,  \antikt\ $R =$ 1.0 jets in $N_{\rm{PV}}$ = 1
  events, corrected to particle level. }
 \label{fig:finalPF10}
 \end{figure}

The planar flow distributions are shown only for events known to be
uncontaminated by pileup, corresponding to events with $N_{\rm{PV}}$ = 1.
These distributions are shown in \figref{finalPF10} for jets reconstructed
with the \antikt algorithm with $R =$ 1.0 for the mass range 130 $< M
<$ 210~GeV, and the corresponding numerical values are given
in \tabref{tablepf10}. The \herwigpp\ 2.4.2 generator predicts
jets with a more planar, isotropic energy distribution than is
observed in data, while version 2.5.1 provides a
very accurate description of the planar flow. The various \pythia\
and \powheg\ Monte Carlo predictions also describe the data well, within uncertainties.

\subsection{\label{sec:respf} Angularity}

\begin{figure}
 \begin{center}
\includegraphics[scale=0.4, angle=0]{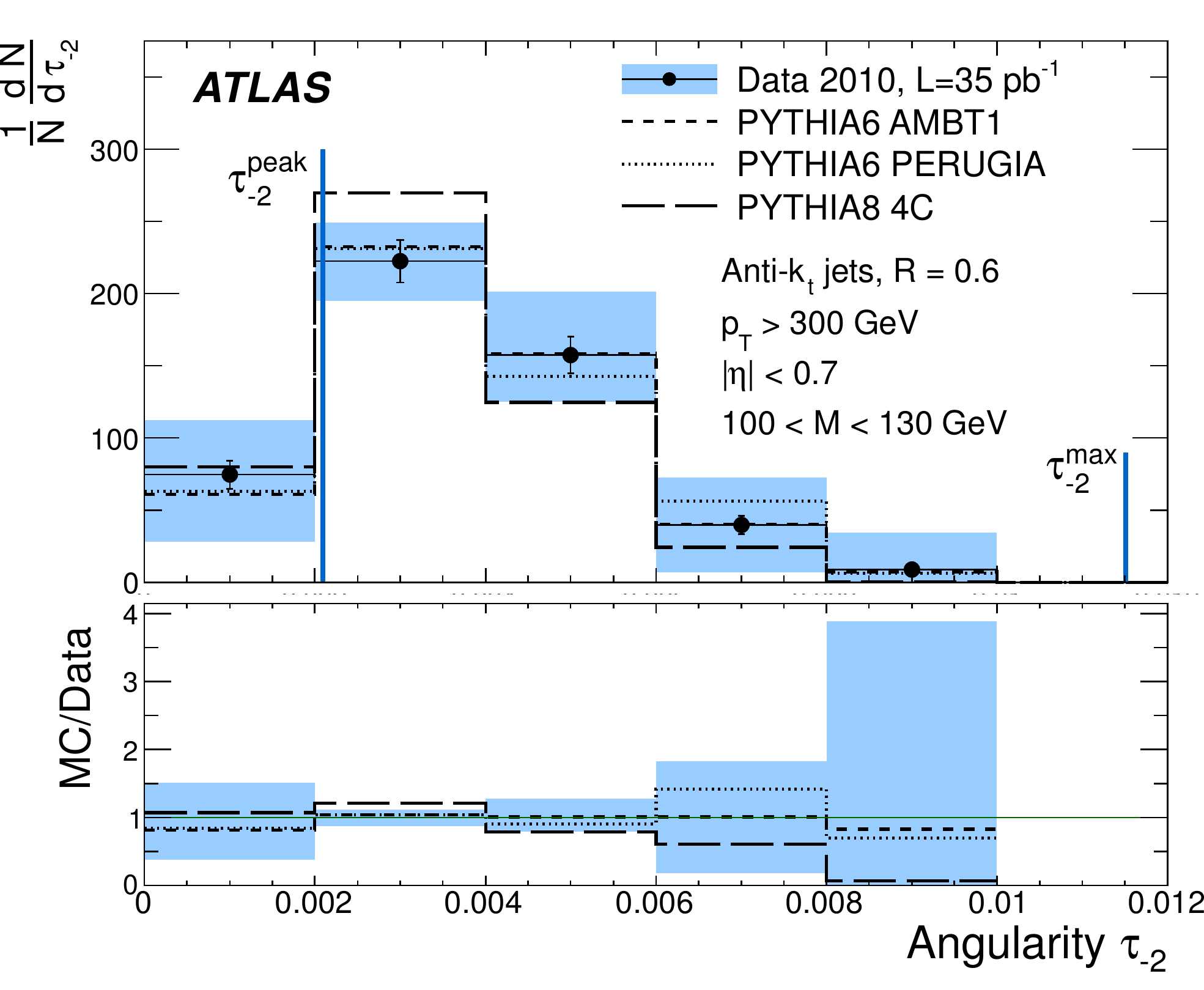}
\includegraphics[scale=0.4,angle=0]{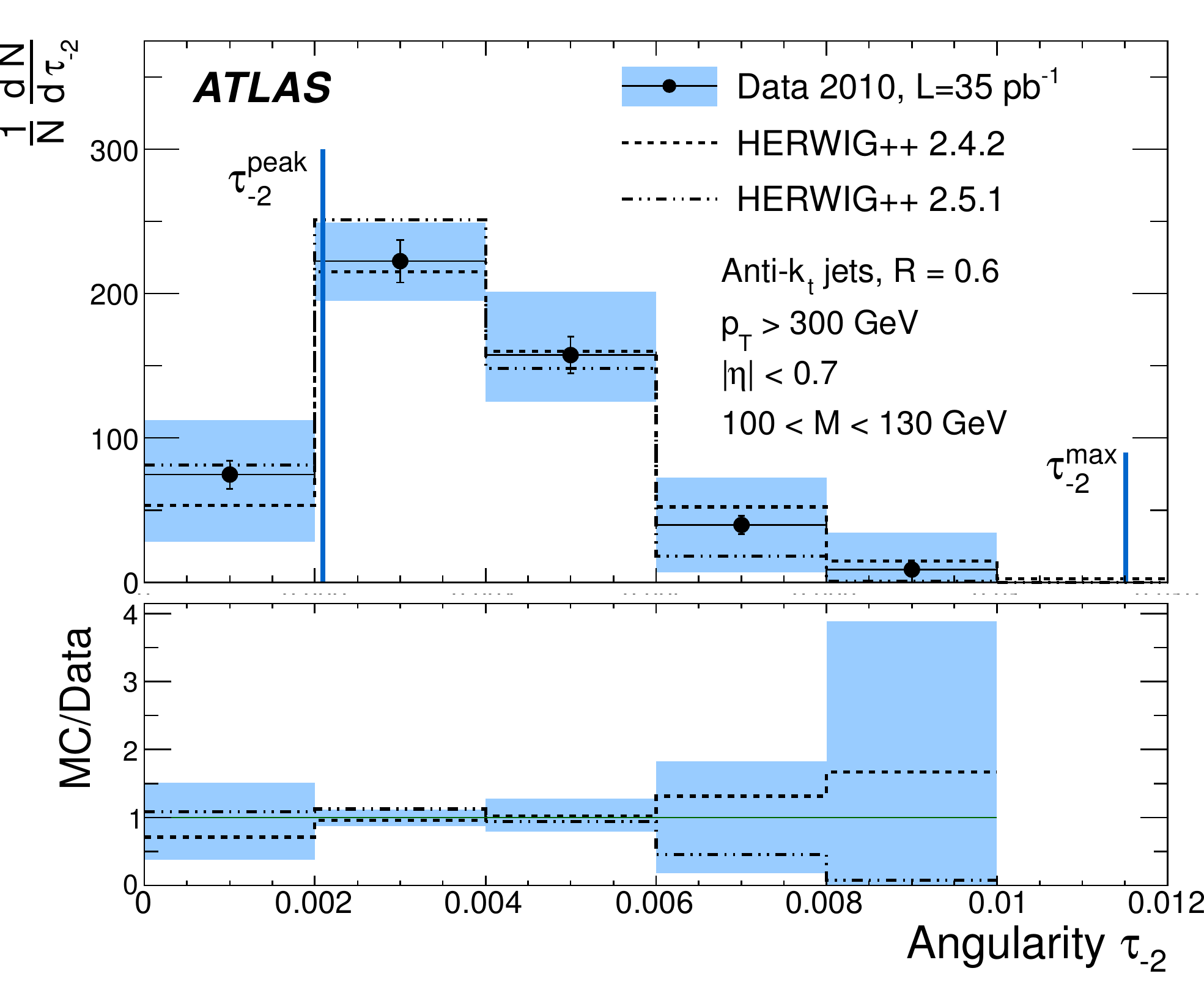}
\includegraphics[scale=0.4, angle=0]{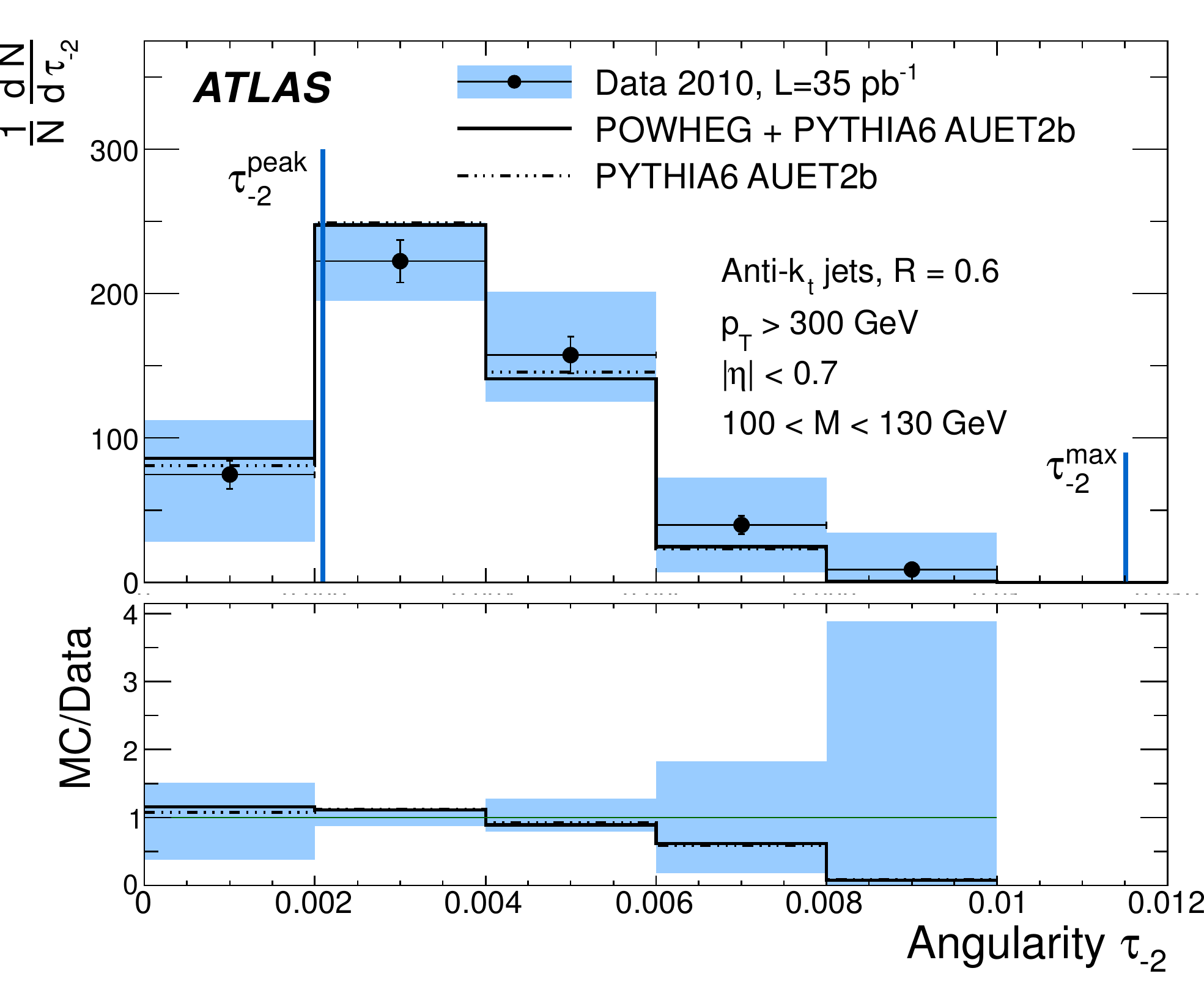}
\end{center}
\caption{The angularity $\tau_{-2}$ distributions
  for leading \pT,  \antikt\ $R =$ 0.6 jets in the mass range 100~$< M <$~130~GeV, in the full 2010
 data set, corrected to particle level. The peak and maximum positions predicted by the small angle
 approximation of \eref{angul2} are indicated. }
\label{fig:finalAng10}
\end{figure}

The $\tau_{-2}$ distribution for \antikt $R =$ 0.6 jets in the mass region 
100 $< M <$ 130~GeV is presented in \figref{finalAng10}, and the corresponding numerical values are given
in \tabref{tablea6}. The QCD
predictions for the peak position and the maximum value of
$\tau_{-2}$~\cite{CDFmass}, calculated using the averages $\avg{M} =$ 111~GeV
and $\avg{p_{\mathrm{T}}} =$ 434~GeV of the jets in this kinematic
region, are also shown on the distributions. Good agreement is
observed between the data and the Monte Carlo simulation for the shape
of the $\tau_{-2}$ distribution. 

The comparison between data and the
analytic QCD prediction is limited by the
intrinsic resolution of the data distribution; however, there is good
agreement between theory and data within these limitations.
The position of the peak of the distribution, $\tau_{-2}^{\mathrm{peak}}$, indicates that
the majority of jets in this data set can be described by a two-body
substructure in a symmetric \pT\ configuration with respect to the jet
axis. No jets are observed above the small-angle kinematic limit, $\tau_{-2}^{\mathrm{max}}$.

\begin{table}
\begin{tabular}{c c l | c c c c c>{\scriptsize}c}  \hline 
     \multicolumn{3}{c|}{Bin [GeV]  \B \T }    &
     $\frac{1}{N}\frac{dN}{dM}$&$ \pm $&$stat. $&$\pm $&$ sys. $&$ (\times 10^{-4})$[$\frac{1}{\mathrm{GeV}}$] \bigstrut  \\   
   \hline     \hline
20&--&40 & $212 $&$ \pm$ &$\hspace{0.1cm} 2 $&$ \pm $&$ 34 $&  \bigstrut\\
40&--&60 & $152$&$ \pm $ &$\hspace{0.1cm} 1 $&$ \pm $&$ 16 $&  \bigstrut\\
60&--&80 & $65$&$ \pm $ &$\hspace{0.1cm} 1 $&$^{+}_{-} $&$^{10}_{11} $&  \bigstrut\\
80&--&110 & $24$&$ \pm$ &$\hspace{0.1cm} 1 $&$ \pm $&$ 4 $&  \bigstrut\\
110&--&140 & $5.0$&$ \pm$ &$\hspace{0.1cm} 0.2$ & $^{+}_{-} $&$^{0.8}_{1.2} $&  \bigstrut\\
    \hline \hline
   \end{tabular}
\caption{Measured values of the \antikt $R =$ 0.6 jet mass distribution given with their statistical and systematic uncertainties.}
\label{tab:tablemass6} 
\end{table}  

\begin{table}[h]
 \begin{tabular}{c c l | c c c c c >{\scriptsize}c}  \hline 
     \multicolumn{3}{c|}{Bin [GeV]  \B \T }    & $\frac{1}{N}\frac{dN}{dM}$&$ \pm $&$stat.$&$\pm $&$ sys. $&$ (\times 10^{-4})$[$\frac{1}{\mathrm{GeV}}$] \bigstrut  \\
   \hline    \hline 
20&--&55 & $69 $&$ \pm $&$ 1 $&$^{+}_{-} $&$ ^{23}_{24} $ & \bigstrut \\
55&--&90 & $122 $&$ \pm $&$ 1 $&$^{+}_{-} $&$ ^{17}_{18} $ & \bigstrut \\
90&--&125 & $56 $&$ \pm $&$ 1 $&$ \pm $&$ 13 $ & \bigstrut \\
125&--&160 & $22.6 $&$ \pm $&$ 0.4 $&$^{+}_{-} $&$ ^{3.2}_{3.4} $ & \bigstrut \\
160&--&200 & $9.0 $&$ \pm $&$ 0.2 $&$^{+}_{-} $&$ ^{2.4}_{2.3} $ & \bigstrut \\
200&--&240 & $4.3 $&$ \pm $&$ 0.2 $&$^{+}_{-} $&$ ^{2.1}_{1.8} $ & \bigstrut \\
    \hline \hline
\end{tabular}
\caption{Measured values of the \antikt $R =$ 1.0 jet mass
  distribution given with their statistical and systematic uncertainties.}
\label{tab:tablemass10}
\end{table}  	

\begin{table}[h]
      \begin{tabular}{r c l | c c c c c >{\scriptsize}c} \hline 
    \multicolumn{3}{c|}{Bin}        &     $\frac{1}{N}\frac{dN}{dW}$&$ \pm $&$ stat.$&$\pm $&$ sys.$&$ (\times 10^{-1})$ \bigstrut  \\
         \hline  \hline
0&--&0.025 & $61.1 $&$ \pm $&$ 1.2 $&$^{+}_{-} $&$ ^{8.2}_{8.5} $ & \bigstrut \\
0.025&--&0.05 & $106 $&$ \pm $&$ 1 $&$^{+}_{-} $&$ ^{12}_{11} $ & \bigstrut \\
0.05&--&0.1 & $55.3 $&$ \pm $&$ 0.4 $&$^{+}_{-} $&$ ^{10}_{9} $ & \bigstrut \\
0.1&--&0.15 & $26.4 $&$ \pm $&$ 0.3 $&$^{+}_{-} $&$ ^{4}_{3} $ & \bigstrut \\
0.15&--&0.2 & $14.0 $&$ \pm $&$ 0.3 $&$ \pm $&$ 2 $ & \bigstrut \\
0.2&--&0.25 & $7.7 $&$ \pm $&$ 0.2 $&$^{+}_{-} $&$ ^{1.3}_{1.2} $ & \bigstrut \\
0.25&--&0.3 & $4.0 $&$ \pm $&$ 0.2 $&$^{+}_{-} $&$ ^{0.9}_{0.7} $ & \bigstrut \\
    \hline \hline
   \end{tabular}
\caption{Measured values of the \antikt $R =$ 0.6 jet width
  distribution given with their statistical and systematic uncertainties.}
\label{tab:tablew6}
\end{table}  

\begin{table}[h]
       \begin{tabular}{r c l | c c c c c >{\scriptsize}c} \hline
   \multicolumn{3}{c|}{ Bin }       &    $\frac{1}{N}\frac{dN}{dW}$&$ \pm $&$ stat.$&$\pm $&$ sys.$&$ (\times 10^{-1})$ \bigstrut  \\
         \hline  \hline
0&--&0.025 & $12.1 $&$ \pm $&$ 0.5 $&$^{+}_{-} $&$ ^{4.8}_{5.0} $ & \bigstrut \\
0.025&--&0.05 & $55.3 $&$ \pm $&$ 0.8 $&$ \pm $&$ 15.0 $ & \bigstrut \\
0.05&--&0.1 & $50.8 $&$ \pm $&$ 0.4 $&$^{+}_{-} $&$ ^{8.2}_{7.5} $ & \bigstrut \\
0.1&--&0.15 & $33.6 $&$ \pm $&$ 0.3 $&$^{+}_{-} $&$ ^{4.9}_{4.5} $ & \bigstrut \\
0.15&--&0.2 & $21.8 $&$ \pm $&$ 0.3 $&$^{+}_{-} $&$ ^{3.3}_{3.0} $ & \bigstrut \\
0.2&--&0.25 & $15.1 $&$ \pm $&$ 0.2 $&$^{+}_{-} $&$ ^{2.1}_{1.9} $ & \bigstrut \\
0.25&--&0.3 & $10.4 $&$ \pm $&$ 0.2 $&$^{+}_{-} $&$ ^{2.4}_{2.2} $ & \bigstrut \\
0.3&--&0.35 & $7.3 $&$ \pm $&$ 0.2 $&$^{+}_{-} $&$ ^{2.2}_{2.1} $ & \bigstrut \\
0.35&--&0.4 & $5.9 $&$ \pm $&$ 0.2 $&$^{+}_{-} $&$ ^{1.6}_{1.4} $ & \bigstrut \\
    \hline \hline
   \end{tabular}
\caption{Measured values of the \antikt $R =$ 1.0 jet width distribution given with their statistical and systematic uncertainties.}
\label{tab:tablew10} 
\end{table}  

\begin{table}[h]
      \begin{tabular}{r c l | c c c c c >{\scriptsize}c} \hline
  \multicolumn{3}{c|}{  Bin }       &     $\frac{1}{N}\frac{dN}{d\mathcal{E}}$&$ \pm $&$ stat.$&$\pm $&$ sys.$&$ (\times 10^{-1})$ \bigstrut  \\
         \hline  \hline
0 &--&0.2 & $0.2 $&$ \pm $&$ 0.1 $&$ \pm $&$ 0.2$ & \bigstrut \\
0.2&--&0.4 & $1.1 $&$ \pm $&$ 0.3 $&$ \pm $&$ 0.6$ & \bigstrut \\
0.4&--&0.6 & $5.1 $&$ \pm $&$ 0.7 $&$^{+}_{-} $&$ ^{1.5}_{1.3} $ & \bigstrut \\
0.6&--&0.8 & $11.0 $&$ \pm $&$ 0.9 $&$^{+}_{-} $&$ ^{1.3}_{1.4} $ & \bigstrut \\
0.8&--&1.0 & $32.9 $&$ \pm $&$ 1.7 $&$^{+}_{-} $&$ ^{2.4}_{2.7} $ & \bigstrut \\
    \hline \hline
   \end{tabular}
\caption{Measured values of the eccentricity distribution for \antikt $R =$ 0.6 jets
  with $M > $~100~GeV, given with their statistical and systematic uncertainties.}
\label{tab:tablee6}
\end{table} 

\begin{table}[h]
       \begin{tabular}{r c l | c c c c c >{\scriptsize}c} \hline
 \multicolumn{3}{c|}{   Bin   }     &    $\frac{1}{N}\frac{dN}{d\mathcal{E}}$&$ \pm $&$ stat.$&$\pm $&$ sys.$&$ (\times 10^{-1})$ \bigstrut  \\
         \hline  \hline
0&--&0.2 & $0.8 $&$ \pm $&$ 0.1 $&$ \pm $&$ 0.3 $ & \bigstrut \\
0.2&--&0.4 & $4.2 $&$ \pm $&$ 0.2 $&$ \pm $&$ 0.9 $ & \bigstrut \\
0.4&--&0.6 & $9.8 $&$ \pm $&$ 0.3 $&$^{+}_{-} $&$ ^{1.2}_{1.4} $ & \bigstrut \\
0.6&--&0.8 & $17.2 $&$ \pm $&$ 0.4 $&$^{+}_{-} $&$ ^{2.0}_{2.2} $ & \bigstrut \\
0.8&--&1.0 & $18.6 $&$ \pm $&$ 0.4 $&$^{+}_{-} $&$ ^{2.8}_{2.7} $ & \bigstrut \\
    \hline \hline
   \end{tabular}
\caption{Measured values of the eccentricity distribution for \antikt $R =$ 1.0 jets
  with $M > $~100~GeV, given with their statistical and systematic uncertainties.}
\label{tab:tablee10}
\end{table}  

\begin{table}[h]
       \begin{tabular}{rcl | c c c c c >{\scriptsize}c} \hline
 \multicolumn{3}{c|}{   Bin }       &    $\frac{1}{N}\frac{dN}{dP}$  &$ \pm $&$ stat.$&$\pm $&$ sys.$&$ (\times 10^{-1})$ \bigstrut  \\
         \hline  \hline
0&--&0.2 & $5.1$&$ \pm $&$ 0.7 $&$ \pm $&$ 1.4$ & \bigstrut \\
0.2&--&0.4 & $13.6 $&$ \pm $&$ 0.9 $&$^{+}_{-} $&$ ^{2.7}_{3.0} $ & \bigstrut \\
0.4&--&0.6 & $13.8 $&$ \pm $&$ 0.9 $&$^{+}_{-} $&$ ^{2.9}_{2.7} $ & \bigstrut \\
0.6&--&0.8 & $9.5 $&$ \pm $&$ 0.8 $&$^{+}_{-} $&$ ^{1.2}_{0.9} $ & \bigstrut \\
0.8&--&1.0 & $8.1 $&$ \pm $&$ 0.7 $&$^{+}_{-} $&$ ^{1.5}_{1.7} $ & \bigstrut \\
    \hline \hline
   \end{tabular}
\caption{Measured values of the planar flow distribution for \antikt $R =$ 1.0  jets in
  $N_{\rm{PV}}$=1 events with 130~$ < M < $~210~GeV, given with their statistical and systematic uncertainties.}
\label{tab:tablepf10}
\end{table}  

\begin{table}[h]
       \begin{tabular}{r c l | c c c c c c} \hline
\multicolumn{3}{c|}{    Bin   }     &  $\frac{1}{N}\frac{dN}{d\tau_{-2}}$&$ \pm $&$ stat.$&$\pm $&$ sys.$ \bigstrut  \\
         \hline  \hline
0&--&0.002 & $75 $&$ \pm $&$ 10 $&$^{+}_{-} $&$ ^{38}_{46} $ & \bigstrut \\
0.002&--&0.004& $223 $&$ \pm $&$ 15 $&$^{+}_{-} $&$ ^{27}_{28} $ & \bigstrut \\
0.004&--&0.006 & $158 $&$ \pm $&$ 13 $&$^{+}_{-} $&$ ^{44}_{33} $ & \bigstrut \\
0.006&--&0.008& $40 $&$ \pm $&$ 6 $&$ \pm $&$ 33 $ & \bigstrut \\
0.008&--&0.010& $9 $&$ \pm $&$ 5 $&$^{+}_{-} $&$ ^{26}_{9} $ & \bigstrut \\
    \hline \hline
   \end{tabular}
\caption{Measured values of the angularity $\tau_{-2}$ distribution for \antikt $R =$
  0.6 jets
  with 100~$ < M < $~130~GeV, given with their statistical and systematic uncertainties.}
\label{tab:tablea6}
\end{table}  

\section{Conclusions\label{sec:conclusions}}

The properties of high \pT\ ($>$ 300~GeV) jets
reconstructed with the \antikt jet algorithm have been
studied in $pp$ collisions at a center-of-mass energy of 7~TeV.
There is good agreement between data and \pythia\ for all observables, and the \powheg+\pythia\
prediction describes the mass distribution well for jets
with $M > $ 20~GeV. \herwigpp\ 2.4.2 predicts jets with
a slightly more isotropic energy flow and higher mass than observed in
data, while \herwigpp\ 2.5.1 predictions are in good agreement with the data. The angularity measurement of high mass jets agrees with the
small-angle QCD approximations.

\section{Acknowledgements\label{sec:acks}} 

We thank CERN for the very successful operation of the LHC, as well as the
support staff from our institutions without whom ATLAS could not be
operated efficiently.

We acknowledge the support of ANPCyT, Argentina; YerPhI, Armenia; ARC,
Australia; BMWF, Austria; ANAS, Azerbaijan; SSTC, Belarus; CNPq and FAPESP,
Brazil; NSERC, NRC and CFI, Canada; CERN; CONICYT, Chile; CAS, MOST and NSFC,
China; COLCIENCIAS, Colombia; MSMT CR, MPO CR and VSC CR, Czech Republic;
DNRF, DNSRC and Lundbeck Foundation, Denmark; EPLANET and ERC, European Union;
IN2P3-CNRS, CEA-DSM/IRFU, France; GNAS, Georgia; BMBF, DFG, HGF, MPG and AvH
Foundation, Germany; GSRT, Greece; ISF, MINERVA, GIF, DIP and Benoziyo Center,
Israel; INFN, Italy; MEXT and JSPS, Japan; CNRST, Morocco; FOM and NWO,
Netherlands; RCN, Norway; MNiSW, Poland; GRICES and FCT, Portugal; MERYS
(MECTS), Romania; MES of Russia and ROSATOM, Russian Federation; JINR; MSTD,
Serbia; MSSR, Slovakia; ARRS and MVZT, Slovenia; DST/NRF, South Africa;
MICINN, Spain; SRC and Wallenberg Foundation, Sweden; SER, SNSF and Cantons of
Bern and Geneva, Switzerland; NSC, Taiwan; TAEK, Turkey; STFC, the Royal
Society and Leverhulme Trust, United Kingdom; DOE and NSF, United States of
America.

The crucial computing support from all WLCG partners is acknowledged
gratefully, in particular from CERN and the ATLAS Tier-1 facilities at
TRIUMF (Canada), NDGF (Denmark, Norway, Sweden), CC-IN2P3 (France),
KIT/GridKA (Germany), INFN-CNAF (Italy), NL-T1 (Netherlands), PIC (Spain),
ASGC (Taiwan), RAL (UK) and BNL (USA) and in the Tier-2 facilities
worldwide.

\bibliography{jetprops}

\clearpage
\input{atlas_authlist}

\end{document}

%% file: atlas_authlist.tex
% ATLAS Collaboration author list for 21-MAR-2012
% Data extracted on 19-May-2012 for paperid 216
\begin{flushleft}
{\Large The ATLAS Collaboration}

\bigskip

G.~Aad$^{\rm 48}$,
B.~Abbott$^{\rm 111}$,
J.~Abdallah$^{\rm 11}$,
S.~Abdel~Khalek$^{\rm 115}$,
A.A.~Abdelalim$^{\rm 49}$,
O.~Abdinov$^{\rm 10}$,
B.~Abi$^{\rm 112}$,
M.~Abolins$^{\rm 88}$,
O.S.~AbouZeid$^{\rm 158}$,
H.~Abramowicz$^{\rm 153}$
% ...
\bigskip

% ...
$^{10}$ Institute of Physics, Azerbaijan Academy of Sciences, Baku, Azerbaijan\\
$^{11}$ Institut de F\'isica d'Altes Energies and Departament de F\'isica de la Universitat Aut\`onoma  de Barcelona and ICREA, Barcelona, Spain\\
% ...
$^{48}$ Fakult\"{a}t f\"{u}r Mathematik und Physik, Albert-Ludwigs-Universit\"{a}t, Freiburg i.Br., Germany\\
$^{49}$ Section de Physique, Universit\'e de Gen\`eve, Geneva, Switzerland\\
% ...
$^{88}$ Department of Physics and Astronomy, Michigan State University, East Lansing MI, United States of America\\
% ...
$^{111}$ Homer L. Dodge Department of Physics and Astronomy, University of Oklahoma, Norman OK, United States of America\\
$^{112}$ Department of Physics, Oklahoma State University, Stillwater OK, United States of America\\
% ...
$^{115}$ LAL, Universit\'e Paris-Sud and CNRS/IN2P3, Orsay, France\\
% ...
$^{153}$ Raymond and Beverly Sackler School of Physics and Astronomy, Tel Aviv University, Tel Aviv, Israel\\
% ...
$^{158}$ Department of Physics, University of Toronto, Toronto ON, Canada\\
% ...
\end{flushleft}